\documentclass[a4paper,twocolumn,superscriptaddress,11pt,accepted=2019-01-10]{quantumarticle}
\pdfoutput=1
\usepackage[utf8]{inputenc}
\usepackage[english]{babel}
\usepackage[T1]{fontenc}
\usepackage{amsmath}
\usepackage{amsfonts}
\usepackage{graphicx,epstopdf,color,verbatim,enumitem}
\DeclareGraphicsExtensions{.png,.pdf,.eps}
\usepackage[numbers,sort&compress]{natbib}

\usepackage[dvipsnames]{xcolor}
\usepackage[bbgreekl]{mathbbol}
\usepackage{mathrsfs}

\usepackage{tikz}
\usepackage{lipsum}
\usepackage[bbgreekl]{mathbbol}
\usepackage{mathrsfs} 
\usepackage{subfigure}

\newcommand{\eins}{\mathbbm{1}}

\newcommand{\mean}[1]{\left\langle #1\right\rangle}

\newcommand{\ket}[1]{\left|#1\right\rangle}

\newcommand{\bra}[1]{\left\langle #1\right|}

\newcommand{\proj}[1]{\ket{#1}\!\!\bra{#1}}

\newcommand\cE{{\mathcal E}}

\newcommand\cF{{\mathcal F}}
\newcommand\bok{\boldsymbol{k}}
\newcommand{\ii}{\mathrm{i}}

\def\k{{\bf k}}

\def\tr{\mathrm{tr}}
\newcommand{\be}{\begin{equation}}
\newcommand{\ee}{\end{equation}}

\begin{document}

\title{All macroscopic quantum states are fragile and hard to prepare}
\author{Andrea L\'opez Incera}
\email{andrea.lopez-incera@uibk.ac.at}
\affiliation{Institut f\"ur Theoretische Physik, Universit\"at Innsbruck, Technikerstra{\ss}e 21a, 6020 Innsbruck, Austria}
\author{Pavel Sekatski}
\email{pavel.sekatski@unibas.ch}
\orcid{0000-0001-8455-020X}
\affiliation{Departement Physik, Universit\"at Basel, Klingelbergstra{\ss}e 82, 4056 Basel, Switzerland}
\author{Wolfgang D\"ur}
\email{wolfgang.duer@uibk.ac.at}
\orcid{0000-0002-0234-7425}
\affiliation{Institut f\"ur Theoretische Physik, Universit\"at Innsbruck, Technikerstra{\ss}e 21a, 6020 Innsbruck, Austria}

\maketitle
\begin{abstract}
We study the effect of local decoherence on arbitrary quantum states. Adapting techniques developed in quantum metrology, we show that the action of generic local noise processes --though arbitrarily small-- always yields a state whose Quantum Fisher Information (QFI) with respect to local observables is linear in system size $N$, independent of the initial state. This implies that all macroscopic quantum states, which are characterized by a QFI that is quadratic in $N$, are fragile under decoherence, and cannot be maintained if the system is not perfectly isolated. We also provide analytical bounds on the effective system size, and show that the effective system size scales as the inverse of the noise parameter $p$ for small $p$ for all the noise channels considered, making it increasingly difficult to generate macroscopic or even mesoscopic quantum states. In turn, we also show that the preparation of a macroscopic quantum state, with respect to a conserved quantity,  requires a device whose QFI is already at least as large as the one of the desired state. Given that the preparation device itself is classical and not a perfectly isolated macroscopic quantum state, the preparation device needs to be quadratically bigger than the macroscopic target state.
\end{abstract}

\section{Introduction}
The last decade has seen increasing efforts to push quantum mechanics to new regimes, and demonstrate quantum features at ever larger scales. There have been significant advances in the control and manipulation of light and matter, and quantum effects have been demonstrated in a variety of systems of up to mesoscopic size (see e.g. \cite{Wineland2013,Eibenberger2013,Kovachy2015,Hosten2016,Wang2016,Vahlbruch2016, Haine}). But will we ever be able to prepare, maintain and confirm truly macroscopic quantum states such as Schr\"odinger's cat? This is a question that is not only of philosophical interest, but may have impact on the planning and the possibility of realizing future experiments.

It is sort of common knowledge that decoherence destroys quantum features, and that macroscopic quantum states are particularly susceptible. However, this statement is mainly restricted to a variety of examples \cite{Caldeira,Yurke,Milburn,Mermin,Peres,Simon,Zurek,Rae,Carlisle}, and only a few general results have been shown, see e.g. \cite{review} for a review. Some macroscopicity measures are intrinsically connected with the stability of states under decoherence. Ref \cite{Duer2002} indeed takes the fragility under decoherence as a measure for macroscopicity. More recent contributions take a different path, and define macroscopicity by other means (see \cite{review} for an overview). However, in some cases it turns out that the very criteria that render a state macroscopic also imply that it is susceptible to noise. One such example is given by the measures of \cite{Sekatski2014} and \cite{Sekatski2014a}, which are based on how good one can distinguish two (or more) states that are in a quantum superposition with a classical detector - or similarly how much information one can extract. Intuitively, if a measurement apparatus can extract information to distinguish between states, so can the environment. Hence there exists a specific decoherence process that destroys quantum coherence and thus macroscopicity for all such states. Similar conclusions were obtained in \cite{korsbakken, park,FrowisNest2013}.  These statements are however restricted to specific decoherence processes.

As argued in \cite{Froewis2012,review}, the variance of a state w.r.t. a local observable $A$ \cite{Shimizu2002} --or more generally the Quantum Fisher Information (QFI), which is the convex roof extension of the variance for mixed states-- can serve as a good measure for macroscopicity. In fact many approaches to quantify macroscopicty make directly or indirectly use of the variance, or the corresponding measures are closely related \cite{review}. For a pure quantum state of $N$ particles of the form $|\psi\rangle \propto (|A\rangle + |D\rangle)$ --where $\ket{A}$ and $\ket{D}$ are classical states--, a large variance of $O(N^2)$ implies that the expectation values for the two states w.r.t. $A$ are macroscopically distinct - i.e. the two states are very different. In fact all separable states have a variance or QFI of $O(N)$, while macroscopic quantum states are characterized by a variance or QFI of $O(N^2)$. The QFI also plays a central role in quantum metrology, and in fact provides bounds like the Cram\'er-Rao bound \cite{CramerBound}, for the achievable precision in estimating a parameter.

Here we show that there are fundamental limitations to maintain and prepare truly macroscopic quantum states based on the QFI. In particular, we find that {\it any} generic local decoherence process that acts  independently on the constituents of the multi-particle quantum state renders a state microscopic as the QFI after the noise process is $O(N)$. This result is obtained by adapting techniques developed in the context of quantum metrology \cite{RafaJan,RafaJanNJP}, where channel extension methods allow one to show that a quantum scaling advantage in parameter estimation disappears if one considers generic noise processes \cite{Sekatski2017,RafalPavel2017} by bounding the attainable QFI for any such quantum channel. Similarly, we find that any initial state is mapped to a state with QFI $O(N)$ w.r.t. any local observable under generic noise processes. The only noise processes for which our methods do not ensure linear scaling of the QFI are some restricted rank 1 channels such as dephasing. Our methods also allow us to obtain upper bounds on the effective size \cite{Duer2002} of the system, where the effective size is given by the minimal size of a quantum system that can show the same features as the system in question, i.e. has the same QFI. We consider different types of noise processes and obtain analytic expressions for the bound on the effective size, that scales inversely proportional to the noise parameter $p$ of the single-qubit noise, $N_{\rm eff} \propto 1/p$, for small noise ($p\ll 1$). This implies that it becomes increasingly difficult to maintain states of larger size.

Furthermore, we also obtain limitations on the preparation process of macroscopic states. We find that, in the typical case where the observable $A$ is the generator of the spatial rotation symmetry and hence a conserved quantity for a closed system, the preparation of a state with a certain QFI requires an apparatus that already needs to have at least the same QFI as the target state.  This follows from the equivalent of the processing inequality for the QFI and the simple fact that, for a closed system, any operation commutes with the symmetry. From this one can conclude that the preparation of a macroscopic quantum state with QFI of $O(N^2)$ requires an apparatus with a QFI that is quadratic in $N$. As the preparation apparatus is usually considered to be classical (and not itself in a quantum superposition state), an apparatus of size $M$ will only have a QFI of $O(M)$, and hence the apparatus needs to be quadratically bigger than the target state, $M=O(N^2)$. One may also use our first result to argue that a preparation apparatus will not be perfectly isolated from the environment, and hence the resulting decoherence implies that the QFI of the apparatus is of $O(M)$. Notice that it was also found that a measurement apparatus that is quadratically larger than the system size is required to successfully confirm the presence of a macroscopic quantum state \cite{Skotiniotis2017}. Our result extends this observation to the preparation process.

The paper is organized as follows. In Sec.~\ref{Background} we introduce the QFI and discuss its role in quantum metrology and macroscopicity. We also review the channel extension method in quantum metrology. In Sec.~\ref{section linear scaling} we adapt the method to the macroscopicity scenario and show that also in this case, one obtains a linear QFI for all full rank and rank 2 channels. We discuss rank 1 channels as well, and show that only rank 1 Pauli channels may leave the QFI of states to be of $O(N^2)$. Using a semidefinite program, we calculate upper bounds on the effective size for finite system sizes, and provide analytic results for several noise channels such as local depolarizing noise or amplitude damping in Sec.~\ref{upper eff size}. Finally, we consider the preparation process in Sec.~\ref{prep} and show that for a closed system the QFI with respect to the total spin can not be increased, hence the effective size that one can prepare is severely limited by the size of the device used to prepare it.

\section{Background and notation}\label{Background}
\subsection{QFI and metrology}
Quantum Fisher Information (QFI) has been widely used in the field of metrology, where some physical parameter of interest $\varphi$ is estimated with some precision $\Delta\varphi$ (where $\Delta^2 \varphi$ denotes the Mean Squared Error of the unbiased estimator), bounded by the so-called Cram\'er-Rao bound \cite{CramerBound},
\begin{equation}
\Delta\varphi\geq\frac{1}{\sqrt{\mathcal{F\,}\left(\rho,\frac{d\rho}{d\varphi}\right)}},
\end{equation}
where $\mathcal{F}\left(\rho,\frac{d\rho}{d\varphi}\right)$ denotes the QFI with respect to the state $\rho$. 

The QFI is mathematically defined as 
\begin{equation}
\mathcal{F}(\rho,\dot{\rho})=2\underset{i,j:\lambda_{i}+\lambda_{j}\neq0}{\sum}\frac{|\bra{i}\dot{\rho}\ket{j}|^{2}}{\lambda_{i}+\lambda_{j}},
\end{equation}
where $\rho=\underset{i}{\sum}\lambda_{i}\ket{i}\bra{i}$ is the spectral decomposition of the state $\rho$, and $\dot{\rho}$ denotes the derivative with respect to the parameter $\varphi$, which defines a one-dimensional parametrization of the space of density matrices. In the context of metrology and macroscopicity, the unitary parametrization $\rho(\varphi)=e^{-iH\varphi}\rho e^{iH\varphi}$ is normally used, where $H$ denotes the time-independent hamiltonian that generates the evolution of the state $\rho$.

The general quantum metrology scenario consists of $N$ probes, e.g. atoms or photons, that sense the parameter $\varphi$, so that by measuring the final state one can estimate the value of the parameter. It has been shown that, if the $N$ probes sense the parameter independently, the achievable precision follows the standard scaling ($1/\sqrt{N}$). This scaling can be improved due to quantum properties such as entanglement, leading to the so-called Heisenberg scaling ($1/N$) \cite{Gio2004,Gio2006}, which corresponds to a quadratic scaling ($O(N^2)$) of the QFI with respect to the system size $N$. However, this quadratic enhancement of precision cannot be achieved if decoherence is taken into account \cite{Shaji2007,FujimaraImai,Escher2011}. In the context of metrology, decoherence arises due to a noisy evolution of the probes. In particular, we focus on the simplified model analysed in \cite{RafaJan}, where $N$ parallel, entangled probes sense the parameter. However, the sensing process may be subject to local noise on each probe, so that the global evolution of the state of $N$ probes can be described by the tensor product of the individual quantum channels $\Lambda_{\varphi}^{\otimes N}$. In this model, each probe undergoes a noisy evolution described by a quantum operation that is of the form: $\Lambda_{\varphi}^{met}(\rho)=\mathcal{E\,}(u_{\varphi}\,\rho\,u_{\varphi}^{\dagger})$. The evolution $u_{\varphi}$ is applied as a unitary rotation that encodes the parameter $\varphi$. The decoherence process is introduced as a quantum channel $\mathcal{E}$ characterized by the Kraus operators $k_{i}$, which do not depend on the parameter $\varphi$. Therefore, the complete quantum channel reads
\begin{equation}
\Lambda_{\varphi}^{met}(\rho)=\sum_{i}k_{i}\,u_{\varphi}\,\rho\,u_{\varphi}^{\dagger}\,k_{i}^{\dagger}=\sum_{i}K_{i}^{met}(\varphi)\,\rho\,K_{i}^{met}{}^{\dagger}(\varphi).
\end{equation}
\subsection{Macroscopicity scenario}
In the context of macroscopicity, we are interested in the study of macroscopic quantum states themselves, and how the interaction with their surroundings eventually transforms them into classical systems. The main difference from the metrology scenario is that the focus of study is not the system evolution and how the noise interferes in the sensing of a parameter. In fact, we optimize over all the possible local observables. In our scenario, we use the scaling of the QFI as a criteria to determine the macroscopicity of a state after a given noise process $\Lambda(\rho)$. If the QFI of the state after the noise process still scales as $O(N^2)$, the state has preserved its macroscopicity.

In particular, we define macroscopic quantum states in terms of the effective size \cite{Froewis2012}. A macroscopic state of $N$ particles that has gone through some noise channel, may perform as a quantum state with $N_{\text{eff}}<N$ particles in a metrological task, i.e. the QFI of the state after the noise may scale as a state with $N_{\text{eff}}<N$ particles. The effective size is defined as,
\begin{equation}
N_{\text{eff}}(\rho)=\underset{H:local}{\max}\frac{\mathcal{F}(\rho,H)}{4N},\label{effsize} 
\end{equation}
where $\mathcal{F}(\rho,H)$ denotes the QFI of the state $\rho$ w.r.t a local observable $H$. Therefore, we call a state macroscopic if $N_{\text{eff}}= O(N)$, that is, $\mathcal{F}(\rho,H)=O(N^2)$.

Considering this specific macroscopicity scenario, in which some noise process is applied to the initially macroscopic quantum state, the one-probe quantum channel has the form $\Lambda_{\varphi}^{mac}=u_{\varphi}\mathcal{\,E}(\rho)\,u_{\varphi}^{\dagger}$; where a perfect unitary evolution $u_{\varphi}$ is applied after the noise process $\mathcal{E}(\rho)=\sum_{i}k_{i} \rho k_{i}^{\dagger}$, described by the Kraus operators $k_{i}$. The noise process is applied before since we are interested in computing the QFI of the noisy state to analyse if the state is still macroscopic or not. Note that a perfect evolution is considered, as it does not represent a real physical process, but the "virtual" evolution on which the QFI of the noisy state $\mathcal{E}(\rho)$ depends. This evolution is just used to compute the QFI but, for the purpose of macroscopicity, it does not have any further meaning. The total channel has the form,
\begin{equation}
\Lambda_{\varphi}^{mac}(\rho)=\sum_{i}u_{\varphi}\,k_{i}\,\rho\,k_{i}^{\dagger}\,u_{\varphi}^{\dagger}=\sum_{i}K_{i}^{mac}(\varphi)\,\rho\,K_{i}^{mac}{}^{\dagger}(\varphi),\label{mac K}
\end{equation} 
where $K_{i}^{mac}(\varphi)=u_{\varphi}k_{i}$ are the Kraus operators of the total channel, satisfying $\sum_{i}K_{i}^{mac}{}^{\dagger}(\varphi)K_{i}^{mac}(\varphi)=\eins$. For simplicity, we denote $K_{i}^{mac}(\varphi)$ as $K_{i}(\varphi)$ from now on. This Kraus representation is not unique, as there exist unitary equivalent representations,
\begin{equation}
\tilde{K}_{i}(\varphi)=\sum_{j}U_{\varphi(ij)}\,K_{j}(\varphi),\label{h matrices}
\end{equation} 
where $U_{\varphi}=e^{-i \varphi h}$ denotes a unitary matrix generated by the hermitian matrix $h$.

\subsection{Channel extension method}
Despite the numerous advantages of the QFI, it is not always possible to analytically compute it, especially when considering some scenarios in metrology, where it is also necessary to maximize the QFI over all possible input states to obtain the maximum achievable precision. However, the channel extension method \cite{FujimaraImai} provides an upper bound (CE bound, where CE stands for channel extension) on the QFI of a $N$-probe state, based solely on the Kraus operators that describe the noise channel for one probe. Intuitively, by extending the system space -e.g. by adding an ancilla- and making the channel act trivially on this extension, the precision of an estimator can only increase, i.e.
\begin{equation}
\mathcal{F\,}(\Lambda_{\varphi}(\rho))\leq\max_{\rho_{ext}}\,\mathcal{F\,}(\Lambda_{\varphi}\otimes\eins_{A}\,(\rho_{ext})),
\end{equation}
where $\rho_{ext}\in\mathcal{B}(\mathcal{H}_{S}\otimes\mathcal{H}_{A})$ denotes the joint state system+ancilla. 

Based on this idea, the CE bound can be derived for both one-probe and $N$-probe channels. Considering a setting of $N$ parallel, entangled probes that undergo some local, uncorrelated noise process, the CE bound has the form \cite{FujimaraImai},

\begin{align}
\mathcal{F\,}(\Lambda_{\varphi}^{\otimes N}(\rho^{N}))&\leq\max_{\rho_{ext}^{N}}\,\mathcal{F\,}(\Lambda_{\varphi}^{\otimes N}\otimes\eins_{A}\,(\rho_{ext}^{N}))\\&\leq4\,\min_{\tilde{K}}\,\{N\,||\alpha_{\tilde{K}}||+N\,(N-1)\,||\beta_{\tilde{K}}||^{2}\},\label{CE bound} 
\end{align}
where  $\rho_{ext}^{N}\in\mathcal{B}(\mathcal{H}_{S}^{\otimes N}\otimes\mathcal{H}_{A})$ and $||\circ||$ denotes the operator norm. The parameters $\alpha_{\tilde{K}}$ and $\beta_{\tilde{K}}$ are defined in terms of the Kraus operators of the one-probe channel {$\tilde{K}_{i}$} as,

\begin{align}
\alpha_{\tilde{K}}&=\boldsymbol{\dot{\tilde{K}}}^{\dagger}\boldsymbol{\dot{\tilde{K}}}=\sum_{i}\dot{\tilde{K}}_{i}^{\dagger}\dot{\tilde{K}}_{i},\label{alfatilde}\\\beta_{\tilde{K}}&=i\boldsymbol{\dot{\tilde{K}}}^{\dagger}\boldsymbol{\tilde{K}}=i\sum_{i}\dot{\tilde{K}}_{i}^{\dagger}\tilde{K}_{i},\label{alphabetaKtilde} 
\end{align}
where $\dot{\tilde{K_{i}}}(\varphi)=\partial_{\varphi}\tilde{K}_{i}(\varphi)$. Throughout this work, we make use of bold letters to designate the vector character of the Kraus operators, e.g. $\boldsymbol{\tilde{K}}$ denotes a column vector with {$\tilde{K}_{i}$} as its elements. We also remark that the optimization over all possible Kraus representations in eq.~\eqref{CE bound} is equivalent to a minimization over the possible $h$ matrices (see eq.~\eqref{h matrices}).

This CE bound can be directly applied to the macroscopicity scenario, the only difference being the definition of the Kraus operators, that in this case have the form $K_{i}(\varphi)=u_{\varphi}k_{i}$ (note that the evolution is applied after the noise process).

\section{Linear scaling of QFI}\label{section linear scaling}

In this section, we apply the methods used in the context of metrology \cite{RafaJan, RafaJanNJP} to macroscopicity. In particular, we study the scaling of the QFI bound \eqref{CE bound} as a way to quantify the macroscopicity of states that have gone through a decoherence process. We show that all noise processes, except some restricted rank 1 Pauli channels, destroy the macroscopicity of the initial quantum state, as the QFI bound of the decohered state scales at most linearly in $N$. A sufficient condition for the bound (\ref{CE bound}) to scale linearly in $N$ is $\beta_{\tilde{K}}=0$, i.e. any noise process that satisfies this condition yields a non-macroscopic state.

In order to study the bound \eqref{CE bound} analytically, the expressions for $\alpha_{\tilde{K}}$ and $\beta_{\tilde{K}}$ in eqs.~\eqref{alfatilde} and \eqref{alphabetaKtilde} need to be written in terms of the variables of interest, i.e. the $h$ matrix and the Kraus operators that describe the given noise process {$k_{i}$}.

Considering $N$-qubit states, the one-qubit evolution $u_{\varphi}=e^{-i \varphi H}$ can be described as a generic rotation in the Bloch sphere, i.e. $H=\sigma_{\vec{n}}=\vec{n}\cdot\vec{\sigma}=n_{1}\sigma_{1}+n_{2}\sigma_{2}+n_{3}\sigma_{3}$. Given that $\boldsymbol{\tilde{K}}=U_{\varphi}u_{\varphi}\boldsymbol{k}$ (see eq.~\eqref{mac K} and \eqref{h matrices}) -where $U_{\varphi}$ acts on the space of Kraus operators and $u_{\varphi}$ acts on the Hilbert space of the one-qubit system-, its derivative is $\boldsymbol{\dot{\tilde{K}}}=-i(h+H)U_{\varphi}u_{\varphi}\boldsymbol{k}$. Hence, the expressions \eqref{alfatilde}, \eqref{alphabetaKtilde} for $\alpha_{\tilde{K}}$ and $\beta_{\tilde{K}}$ have the form,

\begin{align}
\alpha&=\boldsymbol{k^{\dagger}}h^{2}\boldsymbol{k}+\boldsymbol{k^{\dagger}}H^{2}\boldsymbol{k}+2\,\boldsymbol{k^{\dagger}}hH\boldsymbol{k},\label{alphaguay}\\\beta&=-\boldsymbol{k^{\dagger}}(h+H)\boldsymbol{k}.\label{alphabeta} 
\end{align}
Note that we denote $\alpha_{\tilde{K}}$, $\beta_{\tilde{K}}$ as $\alpha$, $\beta$ for simplicity. We remark that, for the purpose of this work, the free variables --over which we optimize-- are the Gauge hamiltonian $h$, that comes from the unitary equivalence of Kraus representations; and the physical hamiltonian $H$, that is also optimized in the context of macroscopicity, where we do not deal with any specific physical evolution. The Kraus operators $\bold{k}$ are fixed depending on the noise process we study.

In the following subsections, we study how initially macroscopic quantum states are affected by different noise processes. Thus, we start by analysing if such processes fulfill the condition $\beta=0$ ($QFI=O(N)$), which, in turn, means that they destroy the macroscopicity of the initial states.

\subsection{Linear scaling of the QFI bound for general channels}
We consider general full rank, rank 2 and rank 1 channels separately to show that all noise processes satisfy the condition $\beta=0$ in eq.~\eqref{alphabeta}, except for rank 1 Pauli channels. Thus, the macroscopicity of the initial state cannot be maintained after such noise processes.

For the purpose of this section, it is enough to show that the different channels can fulfill $\beta=0$. Thus, we only need to demonstrate that the term $\boldsymbol{k^{\dagger}}h\boldsymbol{k}$ in eq.~\eqref{alphabeta} spans the whole vector space of Hermitian operators $\mathcal{S}\equiv\text{span}_h\,   \bok^\dag h \bok = \text{span}\{ \eins, \sigma_x, \sigma_y, \sigma_z\}$; so that the hermitian term $\boldsymbol{k^{\dagger}}H\boldsymbol{k}$ can always be cancelled out (for more details see Appendix~\ref{general observations}).

The specific cases are analysed in detail in Appendix~\ref{general rank 1} (general rank 1 channels), App.~\ref{general rank 2} (general rank 2 channels) and App.~\ref{general full} (general full rank channels). We obtain that all the noise processes can satisfy condition $\beta=0$ except for rank 1 Pauli channels, which implies that their QFI bound is linear in $N$. Hence, the macroscopicity of the initial state can survive, in principle, only if it undergoes a noise process described by a rank 1 Pauli channel.

One possible physical interpretation of this result is the following: if one considers the channel extension method, in which the channels acts trivially on some ancillas, then it is shown in \cite{Duer2014} that there exists an error correction code that can remove rank 1 Pauli noise. Thus, the macroscopicity of the initial state can be maintained after this type of channel because the noise process can be corrected without destroying the state. In the case where no ancillas are allowed, this more physical interpretation is still an open question since, to our best knowledge, there is not an error correction in such scenarios.

For the rest of channels, we can further analyse bound \eqref{CE bound} to get a bound on the effective size (see Sec.~\ref{upper eff size}).

\subsection{Full rank and rank 2 Pauli channels}
In this section, we explicitly compute the conditions $h$ should satisfy to fulfill $\beta=0$ for some useful examples such as full rank and rank 2 Pauli channels. Pauli channels are described as,
\begin{equation}
\mathcal{E}(\rho)=\sum_{i=0}^{r}p_{i}\sigma_{i}\rho\sigma_{i},
\end{equation}
with Kraus operators $\{k_{i}=\sqrt{p_{i}}\sigma_{i}\}_{i=0,..,r}$, that is, each transformation described by $\sigma_{i}$ is applied with probability $p_{i}>0$ (thus, $\sum_{i}p_{i}=1$).

Let us first consider Full rank Pauli channels ($r=3$). We introduce the following notation for the vector $\boldsymbol{k}$ and the $h$ matrix,
\begin{equation}
\boldsymbol{k}=\left(\begin{array}{c}
k_{0}\\
k_{1}\\
k_{2}\\
k_{3}
\end{array}\right)=\left(\begin{array}{c}
\sqrt{p_{0}}\sigma_{0}\\
\sqrt{p_{1}}\sigma_{1}\\
\sqrt{p_{2}}\sigma_{2}\\
\sqrt{p_{3}}\sigma_{3}
\end{array}\right)=\left(\begin{array}{c}
\sqrt{p_{0}}\,\eins\\
\hline \\
\vec{\sigma}_{p}\\
\\
\end{array}\right),
\end{equation}
\begin{equation}
h=\left(\begin{array}{c|ccc}
h_{00} &  & \vec{h}^{\dagger}\\
\hline \\
\vec{h} &  & \mathcal{H}\\
\\
\end{array}\right).
\end{equation}
With this notation, the condition $\beta=0$ (eq.~\eqref{alphabeta}) reads,
\begin{equation}
p_{0}\,h_{00}\,\eins+p_{0}\,(\vec{h}^{\dagger}+\vec{h})\,\vec{\sigma}_{p}+\vec{\sigma}_{p}\,\mathcal{H}\,\vec{\sigma}_{p}+p_{0}\,\sigma_{\vec{n}}+\vec{\sigma}_{p}\,\sigma_{\vec{n}}\,\vec{\sigma}_{p}=0.\label{beta0paulis} 
\end{equation}

Specifically, the conditions for $\beta=0$ that restrict the $h$ components are:
\begin{align}
\sum_{i=0}^{3}p_{i}h_{ii}&=0;\\2\,\sqrt{p_{0}p_{i}}\,Re(h_{0i})-2\,\sqrt{p_{j}p_{k}}\,Im(h_{jk})+\nonumber\\+(p_{0}+p_{i}-p_{j}-p_{k})n_{i}&=0,\label{conditionbetazerofullrank}
\end{align}
for $i,j,k=1,2,3$, where $i\neq j\neq k$ and $j<k$. $Re()$ and $Im()$ denote the real and the imaginary part respectively. These conditions can always be fulfilled with the proper choice of $h$, since all equations are linearly independent and the number of variables exceeds the number of equations.

Similarly, we can study rank 2 Pauli channels and get,
\begin{align}\label{betazerorank2}
\sum_{i=0}^2 p_{i}h_{ii}&=0;\\2\,\sqrt{p_{0}p_{1}}\,Re(h_{01})+(p_{0}+p_{1}-p_{2})n_{1}&=0;\\2\,\sqrt{p_{0}p_{2}}\,Re(h_{02})+(p_{0}+p_{2}-p_{1})n_{2}&=0;\\-2\,\sqrt{p_{1}p_{2}}\,Im(h_{12})+(p_{0}-p_{1}-p_{2})n_{3}&=0\label{betazerorank2bis}.
\end{align}

These conditions will be useful in the following sections (in particular, in Sec.~\ref{analytical}).

\section{Upper bounds on effective size}\label{upper eff size}
Provided that $\beta=0$, the linear bound on the QFI has the form,
\begin{equation}\label{linear bound}
\mathcal{F\,}(\Lambda_{\varphi}^{\otimes N}(\rho^{N}))\leq 4 N \min_{\tilde{K}} ||\alpha_{\tilde{K}}||,
\end{equation}
where the minimization over the possible Kraus representations is, indeed, a minimization over the possible $h$ matrices \footnote{Although the minimization over $h$ gives a tighter bound, we remark that any choice of $h$ that satisfies $\beta=0$ is also a valid upper bound. In any case, it is not known if the upper bound \eqref{linear bound} is saturable for non-extended channels} that satisfy $\beta=0$. 

In the context of macroscopicity, such a linear bound means that the macroscopicity of the initial $N$-qubit state cannot be maintained whenever the system is subjected to a generic local noise process. However, we can extract more information from bound \eqref{linear bound}. By optimizing over the possible local hamiltonians, a bound on the effective size $N_{\text{eff}}$ (Eq.~\eqref{effsize}) can be derived,
\begin{equation}
N_{\text{eff}}(\Lambda_{\varphi}^{\otimes N}(\rho^{N}))\leq \max_{H:local} ||\alpha||_{min},\label{upper bound eff size}
\end{equation}
where $||\alpha||_{min}=\min_{\tilde{K}} ||\alpha_{\tilde{K}}||$. The bound $N_{\text{eff}}^{max}=\underset{H}\max \, ||\alpha||_{min}$ represents the maximum number of particles among which quantum correlations can resist the decoherence action. As a simple example, consider an initial state of $N=1000$ qubits with $N_{\text{eff}}^{max}=20$ after some noise process that fulfills $\beta=0$. The final state is no longer macroscopic, but quantum coherence can be maintained among groups of maximally $20$ qubits.

\subsection{Analytical results}\label{analytical}
In this section, we derive analytical expressions for the bound on the effective size, and in all cases we find that $N_{\text{eff}}^{max}\propto 1/p$ when the noise parameter $p$ is very low. In addition, we derive analytical bounds for amplitude damping and depolarizing channels $\forall p$.

Throughout this section, we write $\alpha$ in terms of the Pauli basis as $\alpha=\sum_{i=0}^{3}a_{i}\sigma_{i}$, with coefficients $a_{i}\in\mathbbm{R}$ obtained by eq.~\eqref{alphaguay}. With this notation, $\alpha=a_{0}\sigma_{0}+\mathcal{C}\sigma_{\vec{c}}$, where $\mathcal{C}=\sqrt{a_{1}^{2}+a_{2}^{2}+a_{3}^{2}}$ and $\vec{c}=\left(\frac{a_1}{\mathcal{C}},\frac{a_2}{\mathcal{C}},\frac{a_3}{\mathcal{C}}\right)$. Hence, its singular values are $\{\left|a_{0}+\mathcal{C}\right|,\,\left|a_{0}-\mathcal{C}\right|\}$, i.e. $||\alpha||=|a_{0}|+\sqrt{a_{1}^{2}+a_{2}^{2}+a_{3}^{2}}$.

\subsubsection{Amplitude damping}
The amplitude damping channel is described by the Kraus operators,
\begin{equation}\label{ampdamKraus}
k_{0}=\left(\begin{array}{cc}
1 & 0\\
0 & \sqrt{1-p}
\end{array}\right),\,\,\,\,k_{1}=\left(\begin{array}{cc}
0 & \sqrt{p}\\
0 & 0
\end{array}\right),
\end{equation}
where $0\leq p < 1$.

Condition $\beta=0$ leads in this case to a fixed $h$ matrix (no optimization over $h$ can be done) of the form,
\begin{equation}
h=\left(\begin{array}{cc}
-n_{3} & (-n_{1}+in_{2})\sqrt{\frac{1}{p}-1}\\
(-n_{1}-in_{2})\sqrt{\frac{1}{p}-1} & (\frac{2}{p}-3)n_{3}
\end{array}\right),
\end{equation}
where $\vec{n}=(n_{1},n_{2},n_{3})$ denotes the hamiltonian direction.

After substituting in eq.~\eqref{alphaguay}, coefficients $a_{i}$ for $\alpha=\sum_{i}a_{i}\sigma_{i}$ read,
\begin{align}
&a_{0}=\left(\frac{1}{p}-2p\right)n_{3}^{2}+\frac{1}{p}+2p-2,\\
&a_{1(2)}=-2\frac{\sqrt{1-p}}{p}n_{1(2)}n_{3},\\
&a_{3}=\left(2p-\frac{2}{p}\right)n_{3}^{2}-2p+2.
\end{align}

As explained above, the analytic expression for $||\alpha||$ is given by $||\alpha||=|a_{0}|+\sqrt{a_{1}^{2}+a_{2}^{2}+a_{3}^{2}}$. Note that, in this case, the $h$ matrix is fixed, so $||\alpha||$ is directly computed (without minimization).

In order to compute the bound on the effective size $N_{\text{eff}}\leq \underset{\vec{n}}\max ||\alpha||$, we only need to optimize over the hamiltonian directions. Considering that $|\vec{n}|=1$, the optimal $\vec{n}$ is obtained by solving $\frac{\partial ||\alpha||}{\partial n_{3}}=0$. Note that, in this case, the norm depends on $a_1^2+a_2^2$, which in turn gives $n_1^2+n_2^2$ that can be written in terms of $n_3$ as $1-n_3^2$. Since we study the function $||\alpha||$ in the region $n_3\in [-1,1]$, we can distinguish two regimes,
\begin{description}
  \item[$\bullet$ $1/\sqrt{2}<p<1$] There are two maxima at $n_3=\pm \frac{(1-p)\sqrt{1+p}}{\sqrt{p(p^2+p-1)}}$. The bound on the effective size is,
  \begin{equation}
  N_{\text{eff}}\leq \frac{3}{p}-\frac{1}{p^2}+\frac{2-3p}{p^2+p-1}.
\end{equation}    
  \item[$\bullet$ $0\leq p\leq 1/\sqrt{2}$] There are only local maxima at $n_3=\pm 1$, that give the bound,
  \begin{equation}
  N_{\text{eff}}\leq \frac{4}{p}-4.
\end{equation}
\end{description}

\subsubsection{Local depolarizing noise}
Local depolarizing noise is a full rank Pauli channel of the form,
\begin{equation}
\mathcal{E}(\rho)=(1-p)\rho+\frac{p}{3}\,\sum_{i=1}^{3}\sigma_{i}\rho\sigma_{i},
\end{equation}
where $0\leq p<0.75$ and the Kraus operators are $k_{0}=\sqrt{1-p}\sigma_{0}$, $\{k_{i}=\sqrt{\frac{p}{3}}\sigma_{i}\}_{i=1,2,3}$. Due to the symmetry of the channel, $||\alpha||$ does not depend on the direction of the hamiltonian $H$, so w.l.o.g. we can choose $\vec{n}=(0,0,1)$. Thus, the bound on the effective size is directly given by $N_{\text{eff}}\leq ||\alpha||_{min}$.

Making use of the numerical methods described in Sec.~\ref{numerical res}, we obtain that the optimal $h$ reads,
\begin{align}
h_{03}&=h_{30}=Re(h_{03}),\nonumber\\
h_{12}&=h_{21}^{*}=iIm(h_{12}),\nonumber\\
h_{ij}&=0 \,\,\,\text{otherwise},
\end{align} 
whose terms are related by the condition $\beta=0$ (eq.~\eqref{conditionbetazerofullrank}): $Im(h_{12})=\frac{1}{p} \left(\sqrt{3p(1-p)}\,Re(h_{03})+(\frac{3}{2}-2p)\right)$. Considering also that $\vec{n}=(0,0,1)$, coefficients $a_i$ now read,
\begin{align}
a_{0}&=(3-\frac{8}{3}p)\,Re(h_{03})^{2}+\frac{6}{\sqrt{3}}\sqrt{\frac{1}{p}-1}\,Re(h_{03})\nonumber\\&+(\frac{3}{2p}-1),\nonumber\\
a_{1}&=a_{2}=a_{3}=0.
\end{align}

Optimizing the analytical expression $||\alpha||=|a_{0}|$ with respect to $Re(h_{03})$ is straightforward and it leads to the bound
\begin{equation}
N_{\text{eff}}\leq \frac{1}{2p}+\frac{1}{9-8p}-1.
\end{equation}

It can easily be seen that for $p\ll1$, $N_{\text{eff}}\leq \frac{1}{2p}-\frac{8}{9}$.

\subsubsection{Pauli channels with $p\ll 1$}
In this section, we obtain analytical bounds for full rank and rank 2 Pauli channels considering that the amount of noise applied to the macroscopic state is infinitesimal, i.e. the noise parameters $p_i\ll 1$ $\forall i\neq0$.

\textbf{Full rank Pauli channels.} $\mathcal{E}(\rho)=\sum_{i=0}^3 p_i\sigma_i\rho\sigma_i$. For this analysis, $p_0\gg p_i$ and $p_i\ll 1$ for $i=1,2,3$.

First, we consider the expression for $\beta=0$ in eq.~\eqref{conditionbetazerofullrank}. The last term $(p_0+p_i-p_j-p_k)n_i\simeq p_0n_i$ is of order $O(1)$, so that $Re(h_{0i})\sim O(1/\sqrt{p})$ and $Im(h_{jk})\sim O(1/p)$ (where $p$ generically denotes the noise parameters, since $p_1 \sim p_2 \sim p_3$) in order to get $\beta=0$. We remark that we use the index notation $i,j,k=1,2,3$, $i\neq j\neq k$, $j<k$ throughout this section. Furthermore, instead of minimizing over the possible Kraus representations, we make an ansatz for the Gauge hamiltonian $h$ to simplify the analysis. In this case, we consider,
\begin{equation}
h=\left(\begin{array}{c|ccc}
0 &  h_{01} & h_{02} & h_{03}\\
\hline \\
h_{01} &  0 & ih_{12} & ih_{13}\\
h_{02} & -ih_{12} & 0 & ih_{23}\\
h_{03} & -ih_{13} & -ih_{23} & 0\\
\\
\end{array}\right),
\end{equation}
where $h_{ij}\in \mathbbm{R}$  $\forall i,j$ and $h_{0i}\sim O(1/\sqrt{p}),h_{jk}\sim O(1/p)$ as argued above. Considering this, the terms in $h^2$ scale as,
\begin{equation}
h^2\sim\left(\begin{array}{c|ccc}
O(\frac{1}{p}) &  & O(\frac{1}{p\sqrt{p}})\\
\hline \\
O(\frac{1}{p\sqrt{p}}) &  & O(\frac{1}{p^2})\\
\\
\end{array}\right).
\end{equation}

Once we have settled the $h$ matrix, we analyse the resulting $\alpha$ term by term (eq.~\eqref{alphaguay}) in order to get the scaling of the bound $N_{\text{eff}}\leq \underset{H}\max \, ||\alpha||_{min}$. The first term $\boldsymbol{k^{\dagger}}h^{2}\boldsymbol{k}$ scales as $O(1/p)$, whereas the other two terms $2\,\boldsymbol{k^{\dagger}}hH\boldsymbol{k}+\eins$ scale as $O(1)$. Thus, we can approximate $\alpha \approx \boldsymbol{k^{\dagger}}h^{2}\boldsymbol{k}$.

The corresponding $a_i$ coefficients are then,
\begin{align}
&a_0 \simeq \sum_{i=1}^3 h_{0i}^2+\sum_{j<k=1}^3 (p_j+p_k)h_{jk}^2+o(1/p),\nonumber\\
&a_1=a_2=a_3 \simeq 0,
\end{align}
which give $||\alpha||=a_0$.

Taking into account condition \eqref{conditionbetazerofullrank}, i.e. $h_{0i}=\frac{2\sqrt{p_j p_k}h_{jk}-n_i}{2\sqrt{p_i}}$, we analytically minimize $a_0$ over the variables $h_{jk}$, and we get,
\begin{equation}
a_{0_{min}}\simeq \sum_{i=1}^3 \frac{p_j+p_k}{4 p_jp_k+4 p_i(p_j+p_k)} n_i^2,\,\,\,i\neq j\neq k\,\,\,j<k,
\end{equation}
for the optimal value $h_{jk}=\frac{n_i\sqrt{p_jp_k}}{2(p_ip_j+p_ip_k+p_jp_k)}$.

Finally, we analytically maximize over the hamiltonian direction $\vec{n}$ (subject to $|\vec{n}|=1$) and we get,
\begin{equation}
N_{\text{eff}}\leq \frac{p_1+p_2}{4 p_1 p_2+4 p_3(p_1+p_1)}+o(1/p),
\end{equation}
where we have assumed w.l.o.g. that $p_{1,2}>p_3$ \footnote{One can always unitarily rotate the Kraus operators $\bold{k}$ such that the noise parameters are ordered like this.}. The optimal hamiltonian direction is $(0,0,1)$.

It can easily be seen that the previous result for the depolarizing channel ($N_{\text{eff}}^{max}=1/2p+o(1/p)$) is recovered by setting $p_1=p_2=p_3=p/3$.

\textbf{Rank 2 Pauli channels.} $\mathcal{E}(\rho)=\sum_{i=0}^2 p_i\sigma_i\rho\sigma_i$. For this analysis, $p_0\gg p_i$ and $p_i\ll 1$ for $i=1,2$. Equivalently, other combinations of two Pauli matrices can also be considered.

Analogously, we can apply the previous procedure to rank 2 Pauli channels. As before, we can approximate $\alpha \approx \boldsymbol{k^{\dagger}}h^{2}\boldsymbol{k}$ ($O(1/p)$) when we consider the ansatz,
\begin{equation}
h=\left(\begin{array}{c|cc}
0 &  h_{01} & h_{02}\\
\hline \\
h_{01} &  0 & ih_{12}\\
h_{02} & -ih_{12} & 0\\
\\
\end{array}\right),
\end{equation}
where $h_{0i}\sim O(1/\sqrt{p})$ for $i=1,2$ and $h_{12}\sim O(1/p)$. Furthermore, the $\beta=0$ conditions (see Sec.~\ref{section linear scaling}) completely determine $h_{01(2)}\simeq \frac{-n_{1(2)}}{2\sqrt{p_{1(2)}}}$ and $h_{12}\simeq \frac{n_3}{2\sqrt{p_1p_2}}$.

Optimizing over the hamiltonian direction, we get
\begin{equation}
N_{\text{eff}}\leq \frac{1}{4}\left(\frac{1}{p_1}+\frac{1}{p_2}\right)+o(1/p),
\end{equation}
for the optimal direction $\vec{n}=(0,0,1)$.

We have shown that for all the analytical cases studied (full rank and rank 2 Pauli channels and amplitude damping), the effective size bound scales as $N_{\text{eff}}\propto 1/p$ when low values of the noise parameter $p$ are considered. This implies that macroscopic quantum states are extremely fragile under decoherence, since, even if the macroscopic state is subjected to a very small amount of noise, a little increase in the noise parameter leads to a decay in the macroscopicity of the system.

\subsection{Numerical results}\label{numerical res} 
In order to obtain the upper bound \eqref{upper bound eff size}, we first optimize over the Kraus representations numerically. Eq.~\eqref{linear bound} can be transformed into a semi-definite program (SDP) \cite{RafaJan,RafaJanNJP} of the form (see Appendix \ref{semidefinite program}),
\begin{eqnarray}
\text{minimize}&&\,\,\,\,\,t,\nonumber\\\text{subject to}&&\,\,\,\left(\begin{array}{cc}
t\,\eins & \boldsymbol{\dot{\tilde{K}}}^{\dagger}\\
\boldsymbol{\dot{\tilde{K}}} & t\,\eins
\end{array}\right)\succeq0,\\&&\,\,\,\,\beta_{\tilde{K}}=0\nonumber,
\end{eqnarray}
where $||\alpha||_{min}=t_{min}^2$ and $\boldsymbol{\dot{\tilde{K}}}=-i(h+H)\boldsymbol{k}$ (note that we have chosen $U_{\varphi=0}=\eins$ and $u_{\varphi=0}=\eins$ w.l.o.g).

Once we get $||\alpha||_{min}$ via SDP, we optimize over the possible hamiltonians ($H=\sigma_{\vec{n}}$ for qubits) by plotting the parameter $||\alpha||_{min}$ for each hamiltonian direction $\vec{n}$, so that the optimal direction can be easily visualized. We make use of the stereographic coordinates to map each three-dimensional vector $\vec{n}$ inside the Bloch sphere onto a point $P$ inside a circle of radius one. Considering that the QFI is symmetric, we only have to take into account the lower half of the Bloch sphere. Thus, the map is a one-to-one correspondence between $\vec{n}$ and $P$,
\begin{equation}
P=\left(\frac{n_{1}}{1-n_{3}}\,,\,\frac{n_{2}}{1-n_{3}}\right),
\end{equation}
where $\vec{n}=(n_{1,}n_{2},n_{3})$.

We apply these methods to different types of noise processes. The results are presented below.

\subsubsection{Full rank Pauli channels}
In this section, we present the results obtained for different choices of full rank Pauli channels,
\begin{equation}
\mathcal{E}(\rho)=\sum_{i=0}^{3}p_{i}\sigma_{i}\rho\sigma_{i},
\end{equation}
where $\sum_{i}p_{i}=1$.

The effective size bound $N_{\text{eff}}^{max}$ is given by the maximum value of $||\alpha||_{min}$ (yellow points in the figures).

\begin{figure}[ht!]
\centering
\subfigure[$p_{1}=2\cdot10^{-4}$ $p_{2}=10^{-4}$ $p_{3}=4\cdot10^{-4}$\label{fig:rank3 asym tiny}]{\includegraphics[width=0.23\textwidth]{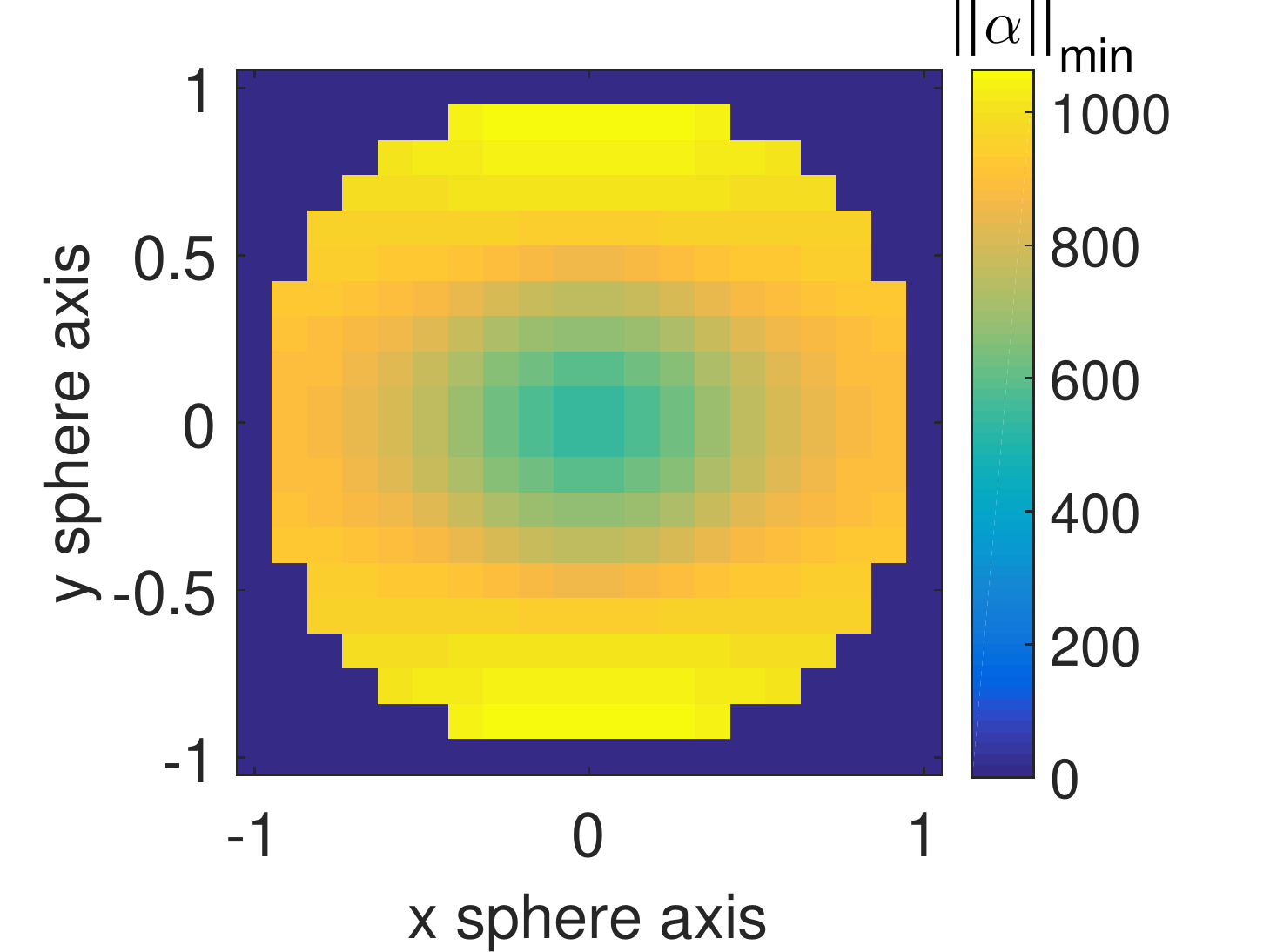}}
\subfigure[$p_{1}=2\cdot10^{-3}$ $p_{2}=10^{-3}$ $p_{3}=4\cdot10^{-3}$\label{fig:rank3 asym small}]{\includegraphics[width=0.23\textwidth]{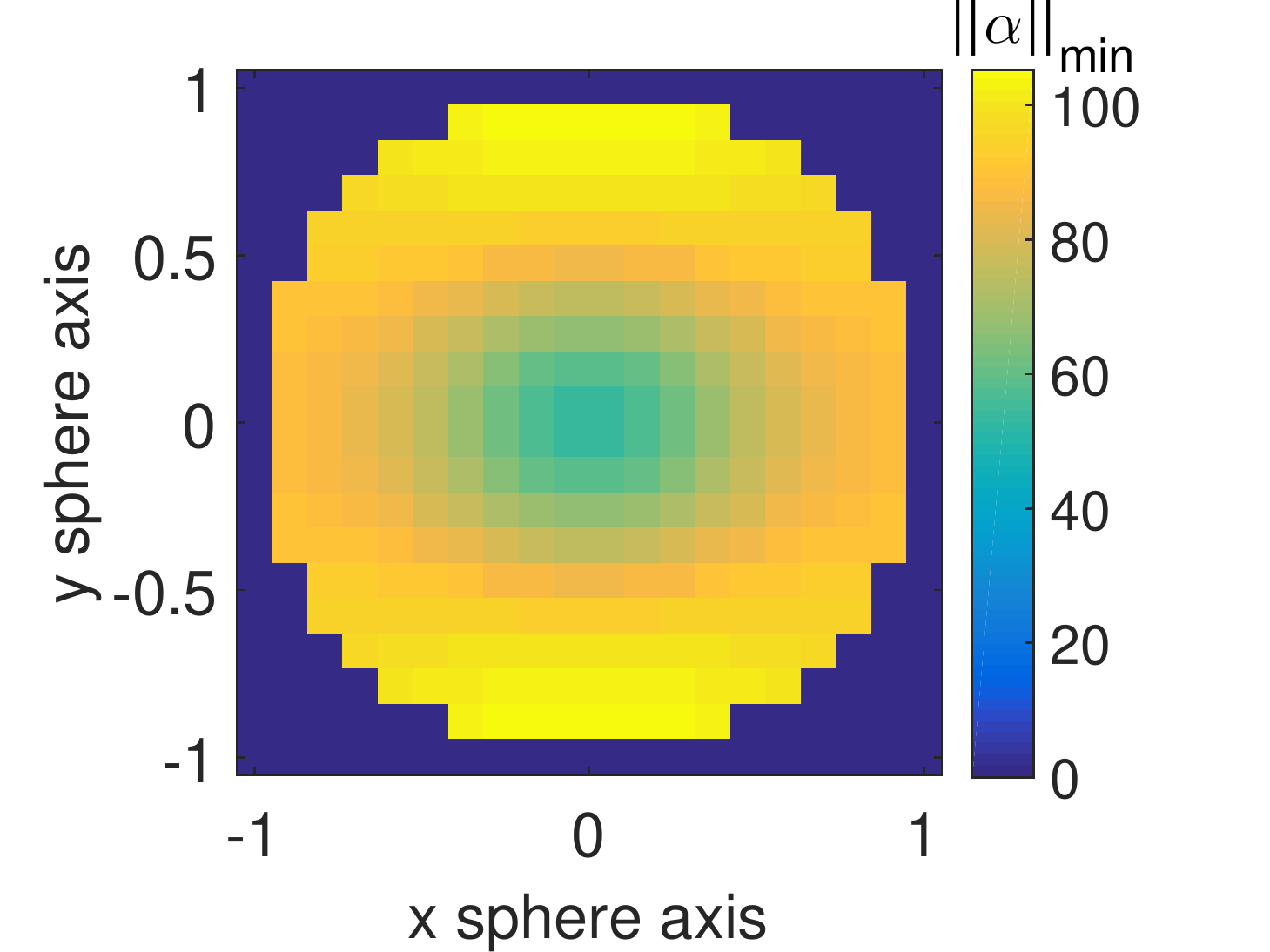}}
\subfigure[$p_{1}=0.02\,\,\,$ $p_{2}=0.01\,\,\,$ $p_{3}=0.04$\label{fig:rank3 asym int}]{\includegraphics[width=0.23\textwidth]{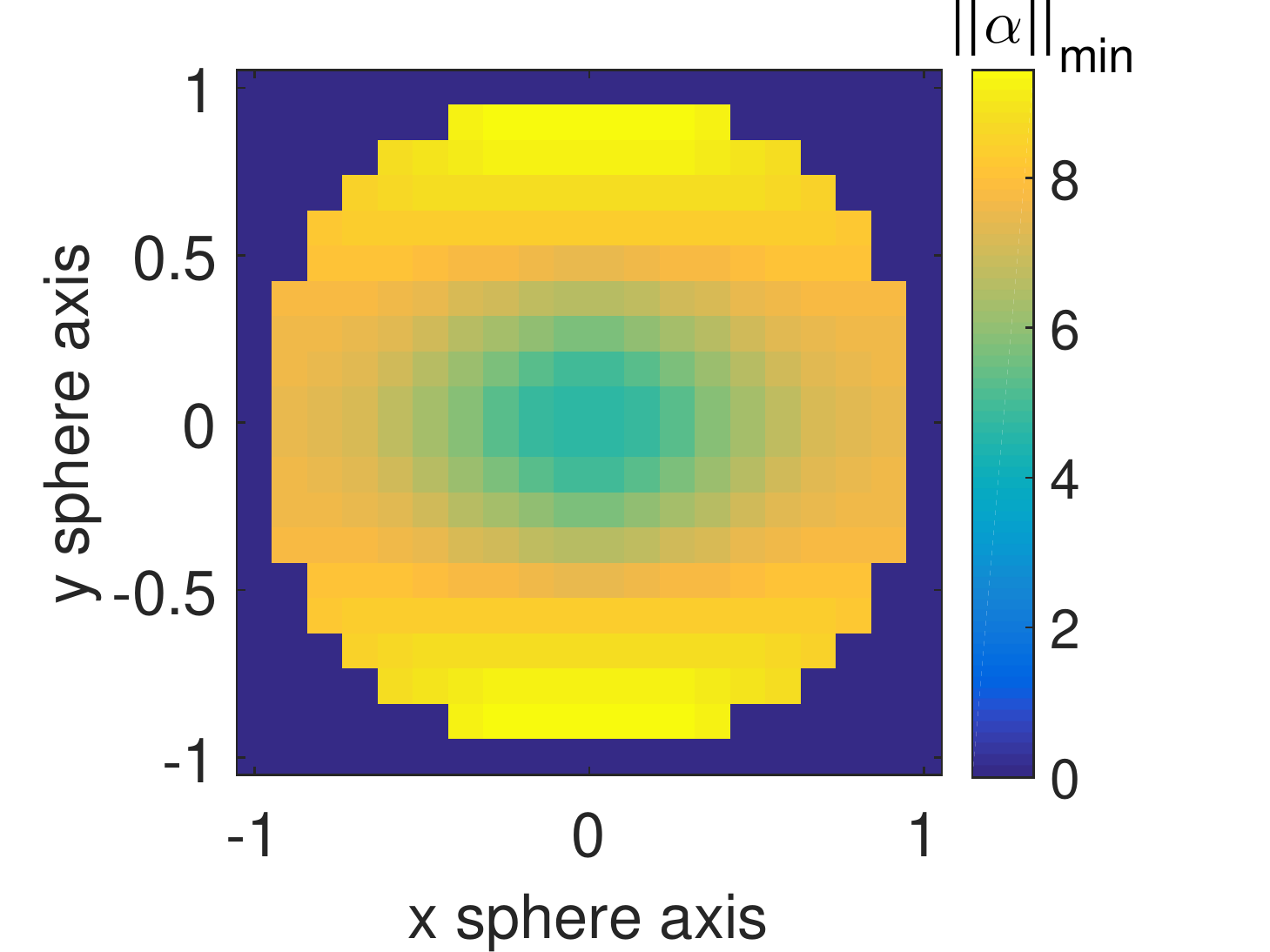}}
\subfigure[$p_{1}=0.2\,\,\,\,\,\,$ $p_{2}=0.1\,\,\,\,\,\,$ $p_{3}=0.4$\label{fig:rank3 high}]{\includegraphics[width=0.23\textwidth]{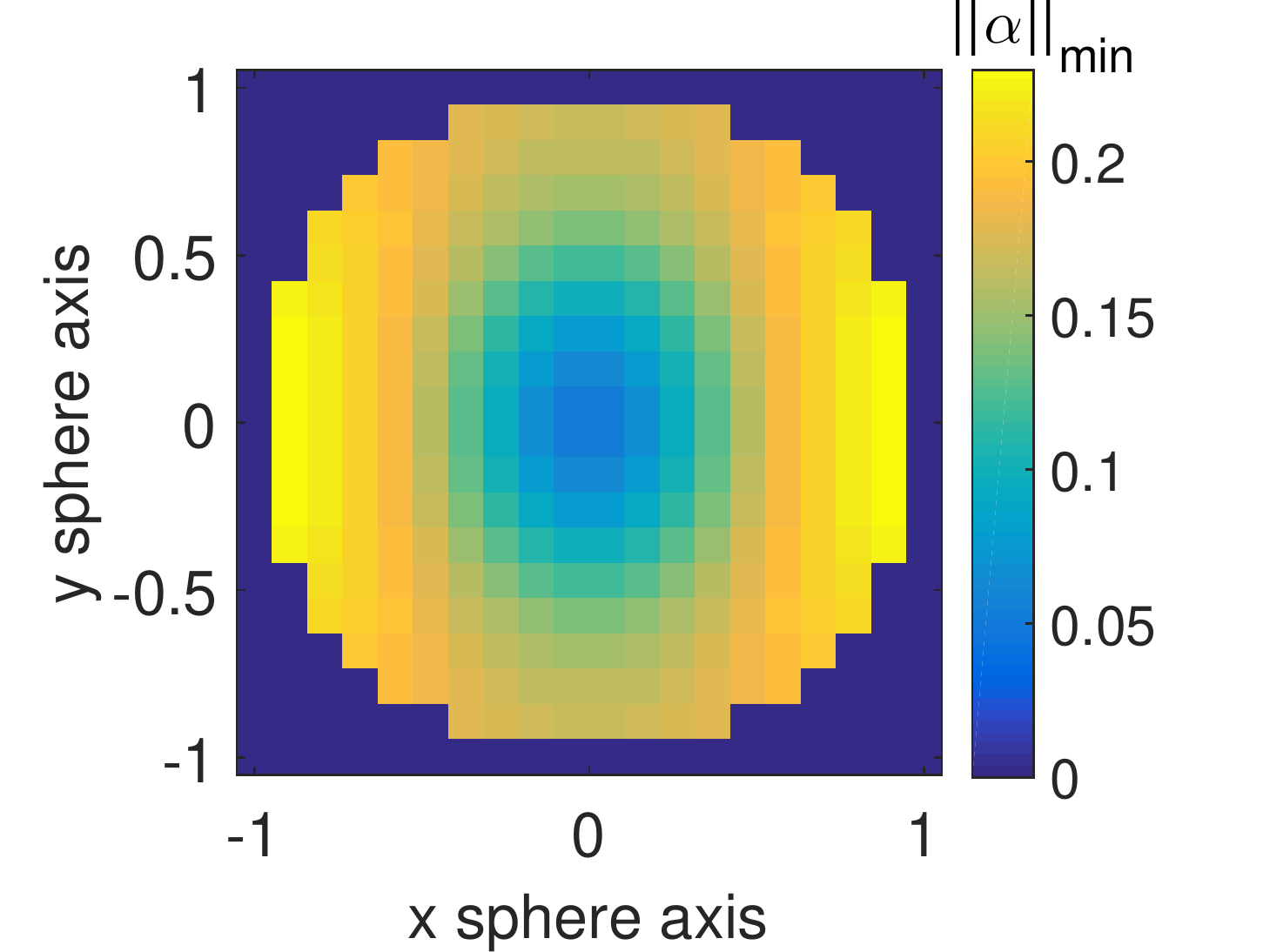}}
\caption{Asymmetric Pauli channel $p_{2}<p_{1}<p_{3}$. $||\alpha||_{min}$ values for the possible hamiltonian directions (points inside the circle) are represented through a color code. The bound on the effective size $N_{\text{eff}}^{max}$ is given by the maximum value of $||\alpha||_{min}$ (yellow points). $\vec{n}_{optimal}$ is close to the $y$ axis in all the cases considered except for fig.~\ref{fig:rank3 high}, where the noise parameters are not infinitesimal, and $\vec{n}_{optimal}$ gets close to the $x$ axis. It can be seen how $N_{\text{eff}}^{max}$ rapidly decays as the noise parameters have higher values. (a)  $N_{\text{eff}}^{max}=1070$ (b) $N_{\text{eff}}^{max}=105$ (c) $N_{\text{eff}}^{max}=9$ (d) $N_{\text{eff}}^{max}=0$. \label{fig:Asymmetric-Pauli-channel}}
\end{figure}

We consider a Pauli channel with $p_{1}\neq p_{2}\neq p_{3}$ (Fig.~\ref{fig:Asymmetric-Pauli-channel}), where $p_{3}>p_{1}>p_2$. Fig.~\ref{fig:rank3 asym tiny} shows the case where a global $0.07\%$ of noise is applied ($p_{0}=0.9993$), leading to an effective size bound $N_{\text{eff}}^{max}=1070$. In Fig.~\ref{fig:rank3 asym small}, all the noise parameters are increased, giving a bound on the effective size of $N_{\text{eff}}^{max}=105$ particles, which means that, with a global $0.7\%$ of noise, quantum coherence is maintained among groups of maximum $105$ particles inside the initial state of $N$ particles. In Fig.~\ref{fig:rank3 asym int}, a global $7\%$ of noise gives $N_{\text{eff}}^{max}=9$. The optimal hamiltonian direction is close to the $y$ axis in these cases. We show in Fig.~\ref{fig:rank3 high} how the optimal hamiltonian direction changes to the $x$ axis when the noise parameters are not infinitesimal.

\subsubsection{Rank 2 Pauli channels}
We now continue with the study of rank 2 Pauli channels, i.e. $\mathcal{E}(\rho)=\sum_{i=0}^{2}p_{i}\sigma_{i}\rho\sigma_{i}$, with $\sum_{i}p_{i}=1$.

As in the previous section, the values of $||\alpha||_{min}$ for each hamiltonian direction $\vec{n}$ are plotted. The maximum value of $||\alpha||_{min}$ (yellow points) is the upper bound on the effective size, $N_{\text{eff}}^{max}$.

\begin{figure}[ht!]
\centering
\subfigure[$p_{1}=2\cdot 10^{-5}$ $p_{2}=0$ $\,\,\,\,\,p_{3}=1.5\cdot 10^{-4}$\label{fig:rank2 tiny}]{\includegraphics[width=0.23\textwidth]{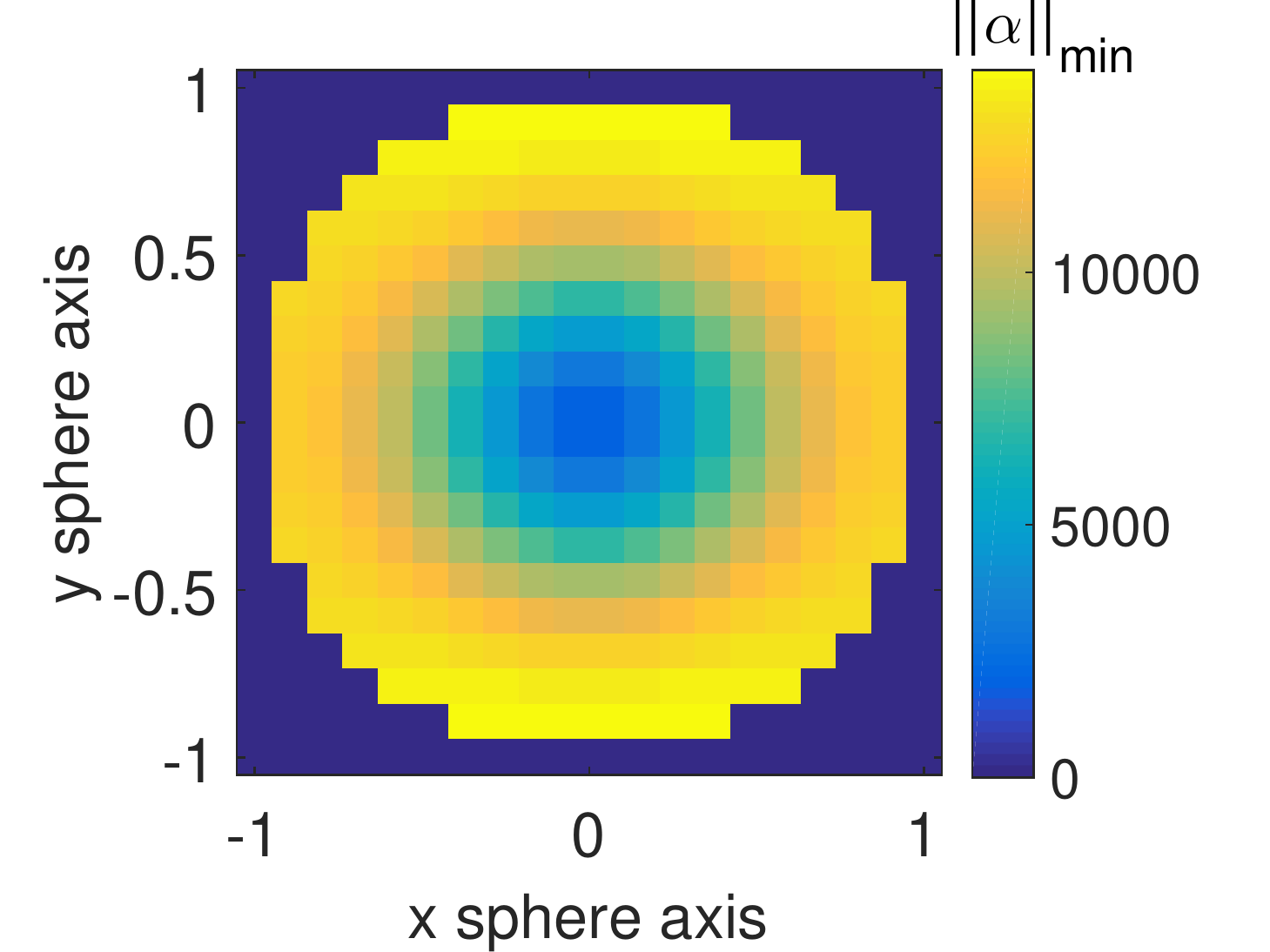}}
\subfigure[$p_{1}=0.002\,\,\,\,$ $p_{2}=0\,\,\,\,\,\,$ $p_{3}=0.015$\label{fig:rank2 small}]{\includegraphics[width=0.23\textwidth]{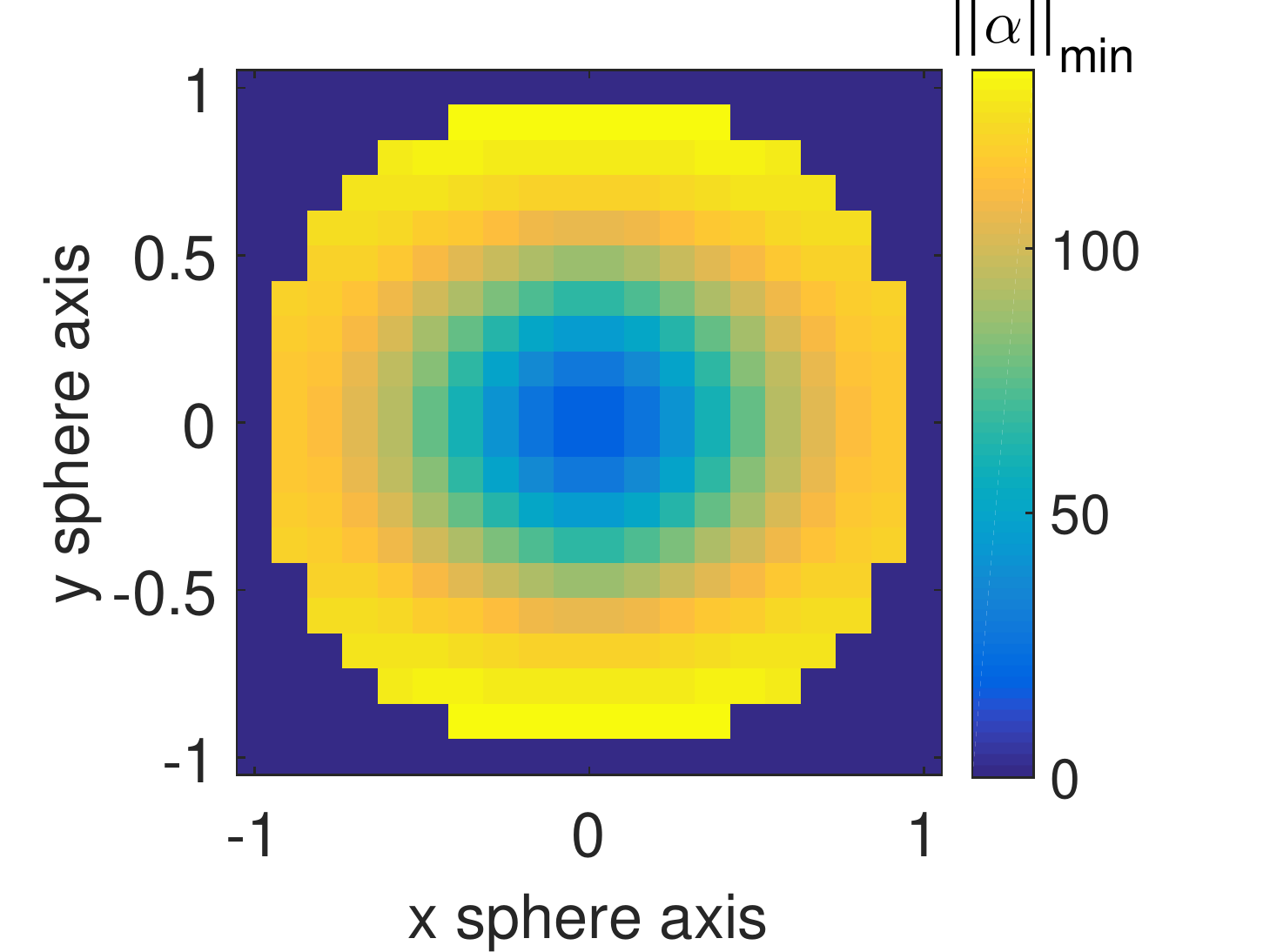}}
\subfigure[$p_{1}=0.02\,\,\,\,\,\,$ $p_{2}=0\,\,\,\,\,\,\,\,\,$ $p_{3}=0.15$\label{fig:rank2 high}]{\includegraphics[width=0.23\textwidth]{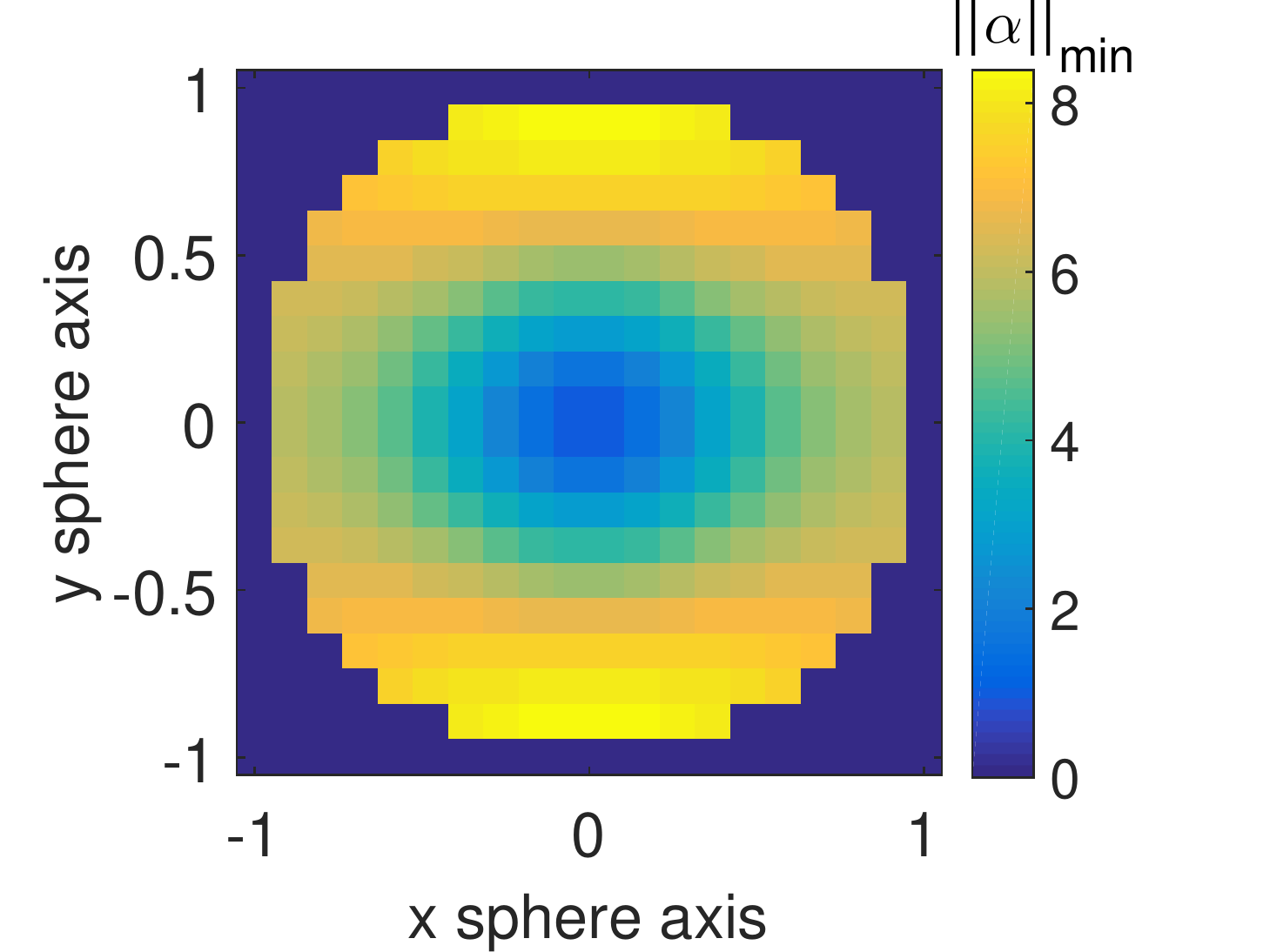}}
\subfigure[$p_{1}=2\cdot 10^{-5}\,\,\,\,\,\,\,\,\,\,\,\,\,\,\,\,\,\,\,\,\,\,\,\,\,\,\,\,\,\,$ $p_{2}=1.5\cdot 10^{-4}$ $p_{3}=0$\label{fig:rank2 noz}]{\includegraphics[width=0.23\textwidth]{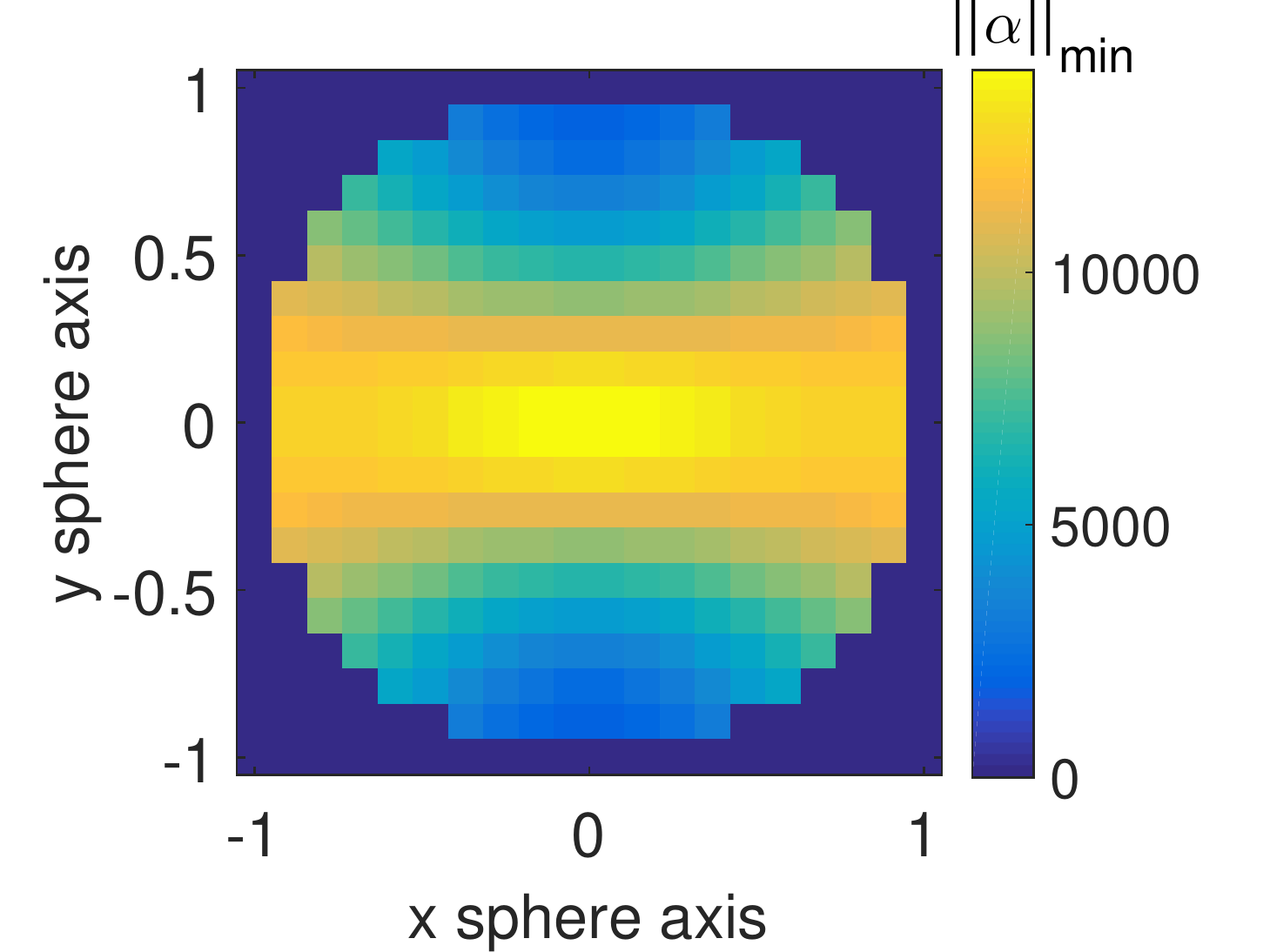}}
\caption{Rank 2 Pauli channel. $||\alpha||_{min}$ values for the possible hamiltonian directions $\vec{n}$ are represented. The maximum value of $||\alpha||_{min}$ (yellow points) corresponds to the bound on the effective size $N_{\text{eff}}^{max}$. $\vec{n}_{optimal}$ is close to the $y$ axis when we consider $p_2=0$ (a,b,c) and it changes to being close to the $z$ axis when $p_3=0$ (d). It can be seen that $N_{\text{eff}}^{max}$ is still high for infinitesimal values of the noise parameters (a,d), but it rapidly decays when the noise parameters are increased (b,c). (a) $N_{\text{eff}}^{max}=14160$ (b) $N_{\text{eff}}^{max}=135$ (c) $N_{\text{eff}}^{max}=8$ (d) Note that this case is analogous to (a), i.e. $N_{\text{eff}}^{max}=14160$, the only difference being the optimal hamiltonian direction.  \label{fig:Rank2-Pauli-channel} }
\end{figure}

We consider Pauli channels with $p_{2}=0$. The effective size bounds are the same (for the same values of the noise parameters) if $p_3=0$ or $p_1=0$ instead. The only difference is the optimal hamiltonian direction (see Fig.~\ref{fig:Rank2-Pauli-channel}). 

In particular, we consider the case where the noise introduced by $\sigma_z$ is much bigger than the $\sigma_x$ ($p_3>p_1$). It can be seen (Fig.~\ref{fig:rank2 tiny}) that, for very low values of the noise parameters ($p_{1}=2\cdot10^{-5},\,\,p_{3}=1.5\cdot 10^{-4}$), the effective size bound is macroscopic ($N_{\text{eff}}^{max}=14160$), corresponding to hamiltonian directions close to the $y$ axis ($0,\pm 1,0$). If noise parameters increase, the effective size bound rapidly decreases to $N_{\text{eff}}^{max}=135$ (Fig.~\ref{fig:rank2 small}) and $N_{\text{eff}}^{max}=8$ (Fig.~\ref{fig:rank2 high}). We show in Fig.~\ref{fig:rank2 noz} how the effective size bound is analogous for rank 2 Pauli channels with $p_3=0$, the only difference being the optimal hamiltonian direction.

\section{Preparation of macroscopic quantum states}\label{prep}

We now step back from the effect of noise on the effective size of quantum states, and consider the limitations on the preparation of macroscopic states in the first place,  arising from the finite size of preparation device and fundamental symmetries of nature. Then we come back to the noise, and interpret the results of the previous section in the new light.

Consider a closed system composed of a "reference frame" $\ket{\text{RF}}$ and a target state $\ket{\psi_0}$, we call the global state $\rho= \proj{\psi_0}\otimes \proj{\text{RF}}$. Both the RF and the target subsystems are composite of $M$ and $N$ particles respectively, that we assume to be qubits for simplicity (this is not crucial for the following argument). Define a local observable $A_\text{tot}$ of the global system as the sum of local observables for the subsystems 
\be
A_\text{tot} = A_\text{sys} + A_\text{RF} = \sum_{i=1}^{N+M} \sigma^{(i)}_{\bf{n}},
\ee
where each $\sigma^{(i)}_{\bf{n}}$ acts on the $i$-th qubit.

As the total system is closed it obeys the usual symmetries and conservation laws. In particular, its time evolution commutes with spatial rotations, as the isotropy of space requires the physics of a closed system to be independent of the choice of orientation of the coordinate system.  Hence, the operations that can be performed on the system also commute with spatial rotations.  The most general operation that can be performed on a quantum system is captured by the notion of quantum instruments: a quantum instrument $\mathscr{E}$ acting on $N+M$ qubits is a collection of CP maps $\cE_k$ such that $\sum_k \cE_k$ is trace preserving and
\be
\mathscr{E}(\rho) = \sum_k \cE_k(\rho) \otimes \proj{k},
\ee
where $\ket{k}$ is a classical label of the outcome of the instrument. As mentioned above, because the system is closed  the isometry of space implies that the rotations around the axis $\bf n$ whose unitary representation is  $U_\varphi= e^{ i \varphi A_\text{tot}}$ commute with the operation $\cE$ 
\be\label{eq:symmetry}
\mathscr{E}\circ U_\varphi = U_\varphi \circ  \mathscr{E},
\ee
{(where $U_\varphi$ is formally extended to act trivially on the outcome label $\ket{k}$).}
Now consider the average Fisher information of the state after the action of the instrument 
\be
\mean{\cF} = \sum_k p_k\, \cF\left( \frac{\cE_k(\rho)}{p_k}, A_\text{tot}\right),
\ee
with $p_k = \tr\, \cE_k(\rho)$ the probability of the outcome $k$. Because all $\ket{k}$ are orthogonal 
we have 
\be
\mean{\cF} = \cF (\mathscr{E}(\rho), A_\text{tot}\otimes\eins_\text{label}). 
\label{eq:QFI instrument}
\ee
This follows from the block diagonal structure of the state $\mathscr{E}(\rho)$ with respect to $k$ and the fact that $ A_\text{tot}\otimes\eins_\text{label}$ only acts non-trivially within each block. The QFI in Eq.~\eqref{eq:QFI instrument} is the susceptibility of the fidelity  
\be
F_\varphi = F (U_\varphi \mathscr{E}\,(\rho)\, U_\varphi^\dag, \mathscr{E}\,(\rho) )
\ee
with respect to small deviations of $\varphi$ from zero $\cF \propto \lim_{\varphi\to 0}\frac{1-F_\varphi}{\varphi^2}$. From the symmetry~\eqref{eq:symmetry} we get 
\be
F_\varphi =  F (\mathscr{E}\,( U_\varphi  \rho \,U_\varphi^\dag)\, , \mathscr{E}\,(\rho) )\geq F ( U_\varphi  \rho \,U_\varphi^\dag\, ,\rho ),
\ee
where for the last step we use the processing inequality for fidelity (the fidelity between two states can not decrease under posprocessing $\mathscr{E}$). From the last inequality we get that the QFI of the initial state can only decrease under manipulations that commute with the observable $A_\text{tot}$
\be
\mean{\cF} \leq \cF(\rho, A_\text{tot}) = \cF(\ket{\psi_0}, A_\text{sys}) +  \cF(\ket{\text{RF}}, A_\text{RF}).
\ee
Note that this relation holds if one traces out some qubits for some outcomes $k$, this follows from the fact that the tracing of subsystems also commutes with $U_\varphi$ which acts locally on each qubit.

To summarise, we have just shown that the QFI of a closed system with respect to an observable $A_\text{tot}$, that commutes with the symmetries e.g. is a generator of the symmetries, can not be increased by any kind of manipulations. This is a direct consequence of the "processing inequality for QFI" -- a direct consequence of the processing inequality for fidelity. 

This limitation on the preparation of macroscopic states allows to  shed new light on the results of previous sections. Typically, in experiments the reference frame system (e.g. a laser beam, a magnet, a double slit...) used for the preparation is not well isolated, but rather an open quantum system subject to all kinds of interactions with the environment and decoherence. Consequently, we know from the results of previous sections that its QFI $\propto O(M)$ can only scale linearly with its size $M$. Hence, in order to prepare a macroscopic state of size $N$, one requires a reference frame system (a preparation device) which is quadratically bigger $M \approx N^2$. We note that similar results have been obtained for  the detection of macroscopic states in \cite{Skotiniotis2017}, where one can also find far-reaching quantitative interpretation for this relation.

\section{Summary and conclusion}
Quantum Fisher Information (QFI) has been used in the context of metrology to derive the Cram\'er-Rao bound for the achievable precision of an estimation protocol. Large scale systems exhibiting quantum properties such as entanglement have been proven to achieve a quadratic scaling of the QFI with the system size $N$, thus increasing the sensitivity of the parameter estimation over the standard scaling (QFI$\sim O(N)$) achieved by separable states. As argued in \cite{Froewis2012}, these different scalings of the QFI can be used to identify genuinely macroscopic quantum states as resource states that improve a metrological protocol.

In this work, we adapt methods used in the field of metrology to study the macroscopicity of $N$-qubit states that have undergone a given noise process. In particular, we analyse the Channel Extension bound (CE bound), that gives an upper bound on the QFI based solely on the Kraus operators of the local noise channel acting on a single qubit.

We show that general full rank and rank 2 channels (including Pauli channels) present a linear scaling of the QFI bound. With respect to rank 1 channels, only rank 1 Pauli channels such as dephasing maintain a quadratic scaling of the QFI bound. This means that initially macroscopic quantum states are fragile under all local, uncorrelated noise channels except for rank 1 Pauli channels.

In addition, we have studied the bound on the effective size, that is, the minimal system size of a quantum system that performs equivalently to the system of study, i.e. the QFI of the state after the noise may scale as a quantum state with $N_{\text{eff}}<N$ particles. Hence, we consider that the state after the noise process is macroscopic if $N_{\text{eff}}=O(N)$.

For all the noise channels considered, the effective size rapidly decays with increasing noise parameter $p$. Only for very low values of the noise parameter ($p\rightarrow0$), $N_{\text{eff}}$ bound can still be of the order of $N$. We obtain analytic expressions for the $N_{\text{eff}}$ bound for depolarizing noise, amplitude damping and full rank and rank 2 Pauli channels with $p\rightarrow0$. In all cases, $N_{\text{eff}}\propto \frac{1}{p}$ in the limit $p\ll 1$.

Finally, we have shown that in a closed system the average QFI of a subsystem with respect to a conserved quantity and after any kind of manipulations can not exceed the original QFI of the total state. Hence the effective size of target state is limited by the QFI of the reference frame state used for the preparation. A preparation device is typically an open system subject to noise, and hence, in lights of the results above, has a QFI scaling linearly with its size. This allows us to conclude that the preparation of a macroscopic state of some size requires a device which is quadratically bigger.

\section*{Acknowledgements} We would like to thank Florian Fr\"owis, Michalis Skotiniotis and Jan Ko{\l}ody\'nski for helpful discussions on parts of this project. This work was supported by the Austrian Science Fund (FWF): P28000-N27, P30937-N27 and SFB F40-FoQus F4012, by the Swiss National Science Foundation (SNSF) through Grant number P300P2\_167749, the Army Research Laboratory Center for Distributed Quantum Information via the project SciNet and the EU via the integrated project SIQS.

\bibliographystyle{unsrtnat}
\bibliography{LinearFisher_macro}

\onecolumn\newpage
\appendix
\section{General Rank 1 noise channels}\label{general rank 1}
General rank 1 noise channels can be generally expressed in terms of the basis $\{\sigma_{i}\}_{i=0...3}$ as,
\begin{align}
k_{0}&=\sum_{i=0}^{3}\mu_{i}\sigma_{i},\\k_{1}&=\sum_{i=0}^{3}\nu_{i}\sigma_{i},\label{kgeneralr1} 
\end{align}
with $\mu_{i},\,\nu_{i}\in\mathbbm{C}\,\,\forall i$. Since $\sum_{i}k_{i}^{\dagger} k_{i}=\eins$, these Kraus operators are related by $k_{0}=\sqrt{\eins-k_{1}^{\dagger}k_{1}}$.

In order to simplify eq.~\eqref{kgeneralr1}, unitarily rotated Kraus operators $\boldsymbol{k^{\prime}}=U \boldsymbol{k}$ can be chosen (e.g. $U=e^{i\theta \sigma_{1}}$, with $\tan{\theta}=i\frac{\nu_{0}}{\mu_{0}}$) so that $k_{1}^{\prime}$ does not contain $\sigma_{0}$. Thus, operator $k_{1}^{\prime}$ (that we rename as $k_{1}$ again for simplicity) can be written as $k_{1}=\vec{\lambda}\cdot\vec{\sigma}$, where $\vec{\lambda}^{\prime}$ denotes the real part and $\vec{\lambda}^{\prime\prime}$ denotes the imaginary part of the complex vector $\vec{\lambda}$. Since we are considering rank 1 channels -$k_{0}$ is completely determined by $k_{1}$-, and the hamiltonian direction is optimized over all possible directions (see eq.~\eqref{effsize}), direction $\vec{\lambda}^{\prime}$ can be arbitrarily chosen and $\vec{\lambda}^{\prime\prime}$ can be decomposed as $\vec{\lambda}^{\prime\prime}=\vec{\lambda}_{\parallel}^{\prime\prime}+\vec{\lambda}_{\perp}^{\prime\prime}$, where $\vec{\lambda}_{\parallel}^{\prime\prime}$ is parallel to $\vec{\lambda}^{\prime}$ and $\vec{\lambda}_{\perp}^{\prime\prime}$ perpendicular. Thus, the expression for $k_{1}$ reads,
\begin{equation}
k_{1}=Re(\lambda_{1})\sigma_{1}+i\,(Im(\lambda_{1})\sigma_{1}+\lambda_{2}\sigma_{2})=\lambda_{1}\sigma_{1}+i\lambda_{2}\sigma_{2},
\end{equation}
where $\lambda_{1}\in\mathbbm{C}$ and $\lambda_{2}\in\mathbbm{R}$. Since $k_{0}=\sqrt{\eins-k_{1}^{\dagger}k_{1}}$,
\begin{equation}
k_{0}=\left(\begin{array}{cc}
\sqrt{A+B} & 0\\
0 & \sqrt{A-B}
\end{array}\right),
\end{equation}
where $A+B>0$, $A-B>0$; with $A=1-|\lambda_{1}|^{2}-\lambda_{2}^{2}$ and $B=2\,Re(\lambda_{1})\lambda_{2}$.

Considering $\beta=-\boldsymbol{k^{\dagger}}(h+H)\boldsymbol{k}$ (eq.~\eqref{alphabeta}), the set of equations for $\beta=0$ is,
\begin{subequations}\label{genr1cond1}
\begin{align}
Ah_{00}+(1-A)h_{11}+2n_{3}B&=0;\\Bh_{00}-Bh_{11}+n_{3}(2A-1)&=0; 
\end{align}
\end{subequations}
\begin{subequations}\label{genr1cond2} 
\begin{align}
\left[Re(\lambda_{1})\,(\sqrt{A+B}+\sqrt{A-B})+\lambda_{2}\,(\sqrt{A+B}-\sqrt{A-B})\right]Re(h_{01})-\nonumber\\
Im(\lambda_{1})\,(\sqrt{A+B}+\sqrt{A-B})\,Im(h_{01})+n_{1}\,(|\lambda_{1}|^{2}-\lambda_{2}^{2}+\sqrt{A^{2}-B^{2}})+2n_{2}Im(\lambda_{1})\lambda_{2}&=0;\\\left[Re(\lambda_{1})\,(\sqrt{A+B}-\sqrt{A-B})+\lambda_{2}\,(\sqrt{A+B}+\sqrt{A-B})\right]Im(h_{01})+\nonumber\\
Im(\lambda_{1})\,(\sqrt{A+B}-\sqrt{A-B})\,Re(h_{01})+n_{2}\,(|\lambda_{1}|^{2}-\lambda_{2}^{2}-\sqrt{A^{2}-B^{2}})-2n_{1}Im(\lambda_{1})\lambda_{2}&=0,
\end{align}
\end{subequations}
where the variables are: $h_{00}$, $h_{11}$, $Re(h_{01})$, $Im(h_{01})$. The set of equations \eqref{genr1cond1} refers only to variables $h_{00}$ and $h_{11}$, whereas the set \eqref{genr1cond2} refers to variables $Re(h_{01})$ and $Im(h_{01})$. We denote them as \textit{set1} and \textit{set2} respectively. In order to show that $\beta=0$, both \textit{set1} and \textit{set2} need to have solutions.

According to Rouch\'e-Capelli theorem \cite{RoucheCapelli}, a system of linear equations $Sx=c$ is consistent (has at least one solution) if and only if the rank of its coefficient matrix ($S$) is equal to the rank of its augmented matrix ($S|c$).

Let us first consider \textit{set1}, with coefficient and augmented matrices,
\begin{equation}
(S_{1}|c_{1})=\left(\begin{array}{cc|c}
A & \,1-A & -2n_{3}B\\
B & -B & n_{3}(1-2A)
\end{array}\right),
\end{equation}
where $\det{S_{1}}=-B$. If $\det{S_{1}}\neq 0$, $rank(S_{1})=rank(S_{1}|c_{1})=2$ so \textit{set1} is consistent. If $\det{S_{1}}=0$, i.e. $B=0$, $rank(S_{1})=1$ so the condition $rank(S_{1}|c_{1})=1$ needs to be fulfilled. This condition is fulfilled if $n_{3}(1-2A)=0$, that is, either $n_{3}=0$ or $A=1/2$.

Similarly, we study \textit{set2}, that has $\det{S_{2}}=2B$. As before, if $B\neq 0$, \textit{set2} is consistent. Otherwise, condition $rank(S_{2}|c_{2})=1$ needs to be fulfilled, together with condition $n_{3}(1-2A)=0$, since both \textit{set1} and \textit{set2} need to be consistent to get $\beta=0$.

Altogether, the options for setting $\beta=0$ are:
\begin{description}
  \item[$\cdot$ $Re(\lambda_{1})\neq 0$, $\lambda_{2}\neq 0$] Both \textit{set1} and \textit{set2} are consistent.
  \item[$\cdot$ (i) $Re(\lambda_{1})=0$, $\lambda_{2}\neq 0$ or (ii) $\lambda_{2}=0$, $Re(\lambda_{1})\neq 0$] In both cases (i) and (ii), rank 1 Pauli channel is recovered, since the Kraus operator $k_{1}$ is of the form 
\begin{equation}
(i) \,\,k_{1}=i|m|\sigma_{\vec{m}},
\end{equation}  
where $\sigma_{\vec{m}}=\vec{m}\cdot \vec{\sigma}$, with $\vec{m}=(\frac{Im(\lambda_{1})}{|m|},\frac{\lambda_{2}}{|m|},0)$ and $|m|=\sqrt{Im(\lambda_{1})^{2}+\lambda_{2}^{2}}$. Or,
\begin{equation}
(ii) \,\,k_{1}=\lambda_{1}\sigma_{1}.
\end{equation}

In this case, the only possibilities for rank 1 Pauli channels to fulfill $\beta=0$ are that, either the hamiltonian direction is the same as the noise direction ($H\propto k_{1}$), or the noise parameters must fulfill the condition $|\lambda_{1}|^{2}+\lambda_{2}^{2}=1/2$ (which is equivalent to a measurement in the eigenbasis of (i)$\sigma_{\vec{m}}$ or (ii)$\sigma_1$). If this is so, the system is consistent (in fact, it has infinite solutions). 
\end{description}

These cases, where $\beta=0$, correspond to types of noise channels that lead to a linear scaling ($O(N)$) of the QFI bound. However, in this context of macroscopicity, we are interested in the effective size, that is, we optimize over all the possible "virtual" hamiltonians $H$. Therefore, the hamiltonian direction can always be chosen to not be parallel to the noise direction.

\section{General considerations on the existence of linear QFI bound}\label{general observations}

The canonical Kraus representation of a general rank-$r$ channel $\cE$ is given by ${\bf k}= (k_0 \, \dots \, k_{r})$, with $\tr\, k_i^\dag k_j = 0$ for $i\neq j$. In order to derive a linear bound on the QFI of state after the noise, we show that there exist choices of Gauge Hamiltonians $h$ which set $\beta=0$ in Eq.~\eqref{alphabeta} for all directions of the physical Hamiltonian $H=\sigma_{\bf n}$. As $\bok^\dag H \bok$ is Hermitian, it is sufficient to show that the Gauge term spans the whole vector space of Hermitian operators
\be
\mathcal{S}\equiv\text{span}_h\,   \bok^\dag h \bok = \text{span}\{ \eins, \sigma_x, \sigma_y, \sigma_z\}.
\ee 

Note that, for our goal, two channels $\cE$ and $\cE'$ related by a posterior unitary transformation $\cE'= U\circ \cE$ are equivalent. This follows from the fact that $U$ can be absorbed in the parametrization of the physical Hamiltonians $H' = U H U^\dag$, and we anyway optimize over all directions of $H$. On the other hand, such a unitary modifies the Kraus operators $k_i \to U k_i$. It follows that we can always choose the unitary transformation such that $k_0 = k_0^\dag$ is Hermitian. Hence, we assume this in the following.

Let us further rewrite the problem in a more convenient form by a reparametrization of $h$. To this end we introduce a positive diagonal matrix 
\be
\text{d} =
\left(
\begin{array}{ccc}
\sqrt{d_0} & & \\
& \ddots& \\
& &\sqrt{d_r}
\end{array}
\right)
\ee
with $d_i = \tr \, k_i^\dag k_i>0$. We have that
\be
\bok^\dag h \bok = \bok^\dag \text{d}^{-1} \text{d}\,  h \,  \text{d} \, \text{d}^{-1}\bok,
\ee
such that a change of variable $h\to \text{d}\, h \, \text{d}$ and $\bok \to \text{d}^{-1} \bok$ does not affect $\mathcal{S}$. The advantage of this parametrization is that the new operators  are orthonormal $\tr\, k_i^\dag k_j =\delta_{ij}$ with respect to the Hilbert-Schmidt inner product. Note that the new operators do not satisfy the trace preserving condition $\sum k_i^\dag k_i \neq \eins$ anymore, instead one has
\be\label{eq:new sum}
\sum d_i\,  k_i^\dag k_i =\eins 
\ee

\section{General Rank 2 noise channels}\label{general rank 2}
  
With the notations of App.~\ref{general observations}, the problem for the general rank-2 noise is to show that $\mathcal{S}$ spans all Hermitian operators for any choice of orthonormal operators $k_0=k_0^\dag, k_1$ and $k_2$.

With the operator $k_0$ being Hermitian, we can choose a basis in which it is diagonal and reads
\be
k_0 = c_\theta \eins + s_\theta \sigma_z
\ee
with $c_\theta>0$ \footnote{In the case where $k_0=\sigma_z$ one can always go to the channel $\sigma_z\circ \cE$, for which $k_0=\eins$. In addition, if $c_\theta < 0$ the Kraus operator can be simply multiplied by $-1$, without affecting the channel $\cE$.}. The first observation is that the diagonal term of the Gauge Hamiltonian $h=\text{diag}(d_0 a, d_1 b, d_2 b)$ yields
\be
\bok^\dag h \bok =
b \eins + d_0(a-b) (\eins + 2 c_\theta s_\theta \sigma_z),
\ee
where we used $ \eqref{eq:new sum}$. If $s_\theta \neq 0$ these terms span all diagonal Hamiltoninas, while for $s_\theta=0$ it only spans $\eins$. We will treat the two cases separately. But before, note that the other two operators $k_1$ and $k_2$ are orthogonal to $k_0$ and hence can be expressed as
\be
k_i = \gamma_i (s_\theta \eins - c_\theta \sigma_z) + \xi_i \sigma_x + \zeta_i \sigma_y
\ee
with $|\gamma_i|^2+|\xi_i|^2+|\zeta_i|^2=1$. In addition, we can choose the free phases $k_i \to e^{i \varphi_i}k_i$ such that $\gamma_i\in \mathbbm{R}$ .

\subsection{$s_\theta\neq 0$ case}

It remains to show that the remaining terms in $h$ allow to span $\sigma_x$ and $\sigma_y$. To do so, first note that there exists a Kraus representation of the channel $\tilde \bok = u^{(12)}(\nu) \bok$, one can also think of it as a unitary transformation of the Gauge Hamiltonian $\tilde h = u^{(1)}(\nu)\, h\, u^{(12)\dag}(\nu)$,
\be
u^{(12)}(\nu)=\left(\begin{array}{ccc}
1&0&0\\
0&c_\nu&s_\nu\\
0&-s_\nu&c_\nu
\end{array}\right)
\ee
after which $\tilde \gamma_2=0$. 

\begin{itemize}

\item
Furthermore, assuming that $\tilde \gamma_1\neq 0$, there exists a subsequent unitary transformation $\bar \bok = u^{(01)}(\nu) \tilde \bok$
\be
u^{(01)}(\nu)=\left(\begin{array}{ccc}
c_\nu&s_\nu&0 \\
-s_\nu&c_\nu& 0\\
0&0&1
\end{array}\right),
\ee
which does not change $\bar k_2=\tilde k_2$ and after which $\bar k_1= \lambda \eins + \mu_x \sigma_x + \mu_y \sigma_y$ has no overlap with $\sigma_z$ and $\lambda \in \mathbbm{R}$. For this new $h$ let us consider the contribution to $\mathcal{S}$ stemming from $\bar h_{12}= r_{12}+\ii\, i_{12}$, one gets
\begin{align}
r_{12}\, (\bar k_1^\dag \bar k_2 +\bar  k_2^\dag \bar  k_1)+\\
i_{12} \, \ii (\bar  k_1^\dag \bar  k_2 - \bar  k_2^\dag \bar  k_1).
\end{align}
Pluging in the expressions for the operators give
\be
(\bar  k_1^\dag\bar  k_2 +\bar  k_2^\dag \bar k_1) = 2 \lambda (\xi_2' \sigma_x +\zeta_2' \sigma_y) + \underbrace{O^r}_{\text{span}(\eins,\sigma_z)}
\ee
\be
\ii(\bar k_1^\dag \bar k_2 -\bar  k_2^\dag \bar k_1) = - 2 \lambda (\xi_2'' \sigma_x +\zeta_2'' \sigma_y) + \underbrace{O^i}_{\text{span}(\eins,\sigma_z)},
\ee
and the contributions $O^{r}$ and $O^{i}$ can be ignored. The operators $(\xi_2' \sigma_x +\zeta_2' \sigma_y)$ and $(\xi_2'' \sigma_x +\zeta_2'' \sigma_y) $ span all off diagonal Hamiltonians (which completes the proof), unless they are linearly dependent. In the latter case we have
\be
\bar k_2 =\tilde k_2 = c_\epsilon \sigma_x +s_\epsilon \sigma_y\propto \xi_2' \sigma_x + \zeta_2' \sigma_y,
\ee 
after we get rid of the phase factor. Considering the terms stemming from $\tilde h_{02}$ with $\tilde k_0=k_0$ gives the following contribution
\be
\ii (\tilde k_0^\dag \tilde k_2 - k_2^\dag k_0)=s_\theta [\sigma_z,  c_\epsilon \sigma_x +s_\epsilon \sigma_y],
\ee
this operator is off-diagonal and orthogonal to $ \xi_2' \sigma_x + \zeta_2' \sigma_y$, hence recover the span of all off-diagonal Hamiltonians.\

\item if $\tilde \gamma_1 = \tilde \gamma_2=0$, we get
\begin{align}
 k_1 = c_\omega \sigma_x + e^{\ii \phi} s_\omega\sigma_y\\
 k_2 = s_\omega \sigma_x - e^{\ii \phi} c_\omega \sigma_y.
\end{align}
Applying the unitary
\be
u 
=\left(\begin{array}{ccc}
1&0&0\\
0&c_\omega&s_\omega\\
0&e^{-\ii \phi}s_\omega &-e^{-\ii \phi}c_\omega
\end{array}\right),
\ee
gives $\binom{\hat k_1}{\hat k_2} = \binom{\sigma_x}{\sigma_y}$. Hence, the following terms
\begin{align}
\frac{\ii}{2}(\hat k_0^\dag \hat k_1 -\hat k_1^\dag \hat k_0 ) = - s_\theta \sigma_y\\
\frac{\ii}{2}(\hat k_0^\dag \hat k_2 -\hat k_1^\dag \hat k_2 ) =  s_\theta \sigma_x
\end{align}
span the off-diagonal Hamiltonians. This completes the proof for the case $s_\theta\neq 0$.
\end{itemize}

\subsection{$s_\theta= 0$ case}

This case corresponds to  $k_0 =\eins$. Without loss of generality we can set the basis of the Pauli operators such that
\be
k_1 = c_\delta \, \sigma_x + \ii s_\delta \, (c_\lambda \sigma_x + s_\lambda \sigma_y)
\ee
with $c_\delta>0$ and $s_\lambda\neq 0$. $k_0$ and $k_1$ yield the following contributions to $\mathcal{S}$
\begin{align}
k_0^\dag k_0 = \eins\\
\frac{1}{2}(k_0^\dag  k_1 + k_1^\dag k_0) = c_\delta \, \sigma_x\\
\frac{\ii}{2}(k_0^\dag  k_1 - k_1^\dag k_0) = -s_\delta (c_\lambda \sigma_x + s_\lambda \sigma_y)\\
k_1^\dag k_1 = \eins + c_\delta s_\delta s_\lambda \sigma_z,
\end{align}
which span the whole vector space unless $c_\delta=1$ and $k_1=\sigma_x$.  We can repeat the same argument for $k_2$, finishing the proof unless it is also a Pauli operator. In consequence, up to a basis change the only remaining case for which the existence of $\beta=0$ has not been demonstrated is $(k_0, k_1, k_2)= (\eins, \sigma_x, \sigma_y)$. But for this channel we obtain
\begin{align}
\frac{1}{2}(k_0^\dag  k_1 + k_1^\dag k_0) =\sigma_x\\
\frac{1}{2}(k_0^\dag  k_2 + k_2^\dag k_0) =\sigma_y\\
\frac{\ii}{2}(k_1^\dag  k_2 - k_2^\dag k_1) =-\sigma_z,
\end{align}
which completes the proof.
 
\section{Full-rank noise}\label{general full}

With the notations of App.~\ref{general observations} the full rank noise is specified by 4 operators
\be
k_i = \xi_i^{k} \sigma_k,
\ee
with orthonormal vectors $\sum_k \xi_i^{k *} \xi_j^k =\delta_{ij}$. So the Kraus operators form an (orthonormal) basis of the complex Hilbert space represented by complex $2\times2$ matrices with the Hilbert-Schmidt inner product. As any two bases are related by a unitary transformation, and full-rank channel in our notation can be related to  
\begin{align}
\tilde k_0 =\eins\\
\tilde k_1 =\sigma_x\\
\tilde k_2 =\sigma_y\\
\tilde k_3 =\sigma_z
\end{align}
by some unitary transformation. In this form, it is direct to see that $\tilde \bok^\dag h \tilde \bok$ spans   all hermitian matrices.

\section{QFI bound as a semi-definite programming} \label{semidefinite program}
Given the matrix $\alpha$, the minimization of its operator norm (maximum singular value) can be expressed as,
\begin{eqnarray}
\text{minimize} &  & \,\,\,\,t,\nonumber \\
\text{subject to} &  & \,\,\,\,t^{2}\,\eins-\boldsymbol{\dot{\tilde{K}}}^{\dagger}\boldsymbol{\dot{\tilde{K}}}\succeq0,\nonumber \\
 &  & \,\,\,\,\beta_{\tilde{K}}=0,\label{eq:SDP NORM form}
\end{eqnarray}
from where it can be seen that $||\alpha||=t^2$.

Due to the Schur complement condition for positive semidefiniteness, eq.~\eqref{eq:SDP NORM form} is equivalent to,
\begin{eqnarray}
\text{minimize} &  & \,\,\,\,t,\nonumber \\
\text{subject to} &  & \,\,\left(\begin{array}{cc}
t\,\eins_{d_1} & \,\,\boldsymbol{\dot{\tilde{K}}}^{\dagger}\\
\boldsymbol{\dot{\tilde{K}}} & \,\,t\,\,\eins_{\small{(r+1)\cdot d_2}}
\end{array}\right)\succeq0,\nonumber \\
 &  & \,\,\,\,\beta_{\tilde{K}}=0,\label{eq: SDP matrix form}
\end{eqnarray}
where the Kraus operators $\{\dot{\tilde{K}}_i\}_{i=0}^{r}$ are $d_2\times d_1$ matrices and $\eins_{d}$ is the $d\times d$ identity matrix. Problem \eqref{eq: SDP matrix form} is a semi-definite program since the Kraus operators $\dot{\tilde{K}}_i=-i\sum_{j}h_{ij}k_{j}-i\sigma_{\vec{n}}k_{i}$ are linear in $h$. Explicitly, condition $\beta_{\tilde{K}}=0$ is in our case $\sum_{ij}h_{ij}k_{i}^{\dagger}k_{j}+\sum_{i}k_{i}^{\dagger}\sigma_{\vec{n}}k_{i}=0$.

To solve problem \eqref{eq: SDP matrix form} we used CVX, a package for specifying and solving convex programs \cite{cvx,gb08}.

\end{document}